\documentclass[%
reprint,
 superscriptaddress,
nofootinbib,
 amsmath,amssymb,
 aps,
 prd,
]{revtex4-2}

\usepackage{graphicx,amssymb,amsmath,amsthm,amsfonts,epsfig,epsf}
\usepackage[linktocpage]{hyperref}
\usepackage[usenames]{color}
\usepackage{epstopdf}
\usepackage{bm}
\usepackage{dcolumn}
\usepackage[latin1]{inputenc}
\usepackage{latexsym}
\usepackage{rotating}
\usepackage{hyperref}
\usepackage{color}
\usepackage{longtable}

\usepackage{enumerate}
\usepackage{tensor,multirow}
\usepackage{url}
\usepackage{tikz,tikz-3dplot}

\usepackage{slashed}

\newcommand{\nn}{\nonumber}

\usepackage[scr=boondox]{mathalfa}

\newcommand{\cm}{\mathscr{m}}

\newcommand{\cq}{\mathscr{q}}
\newcommand{\ca}{\mathscr{a}}

\DeclareFontFamily{U}{min}{}
\DeclareFontShape{U}{min}{m}{n}{<-> udmj30}{}

\begin{document}


\title{Parameterized Post-Einsteinian Framework for Precessing Binaries}

\author{Nicholas Loutrel}
 \email{nicholas.loutrel@uniroma1.it}
 \affiliation{Dipartimento di Fisica, ``Sapienza'' Universit\`a di Roma \& Sezione INFN Roma1,\\ Piazzale Aldo Moro 5, 00185, Roma, Italy}

\author{Paolo Pani}
 \affiliation{Dipartimento di Fisica, ``Sapienza'' Universit\`a di Roma \& Sezione INFN Roma1,\\ Piazzale Aldo Moro 5, 00185, Roma, Italy}

\author{Nicol\'as Yunes}
\affiliation{
 Illinois Center for Advanced Studies of the Universe, Department of Physics\\ University of Illinois at Urbana-Champaign, Champaign, Illinois, USA
}%

\date{\today}

\begin{abstract}

In general relativity, isolated black holes obey the no hair theorems, which fix the multipolar structure of their exterior spacetime. 
However, in modified gravity, or when the compact objects are not black holes, the exterior spacetime may have a different multipolar structure. 
When two black holes are in a binary, this multipolar structure determines the morphology of the dynamics of orbital and spin precession. 
In turn, the precession dynamics imprint onto the gravitational waves emitted by an inspiraling compact binary through specific amplitude and phase modulations. 
The detection and characterization of these amplitude and phase modulations can therefore lead to improved constraints on fundamental physics with gravitational waves. 
Recently, analytic precessing waveforms were calculated in two scenarios: (i) dynamical Chern-Simons gravity, where the no-hair theorems are violated, and (ii) deformed compact objects with generic mass quadrupole moments. 
In this work, we use these two examples to propose an extension of the parameterized post-Einsteinian~(ppE) framework to include precession effects. 
The new framework contains $2n$ ppE parameters $(\mathscr{b}^{\rm ppE}_{(m',n)}, b^{\rm ppE}_{(m',n)})$ for the waveform phase, and $2n$ ppE parameters $(\mathscr{a}^{\rm ppE}_{(m',n)}, a^{\rm ppE}_{(m',n)})$ for the waveform amplitudes. 
The number of ppE corrections $n$ corresponds to the minimum number of harmonics necessary to achieve a given likelihood threshold when comparing the truncated ppE waveform with the exact one,
and $(m',n)$ corresponds to the harmonic numbers of the harmonics containing ppE parameters.
We show explicitly how these ppE parameters map to the specific example waveforms discussed above. 
The proposed ppE framework can serve as a basis for future tests of general relativity with gravitational waves from precessing binaries.
\end{abstract}

\maketitle


\section{\label{sec:intro}Introduction}

Observations of gravitational waves~(GWs) from binary systems seek to answer several important questions, among which are ``what is the nature of the compact objects comprising these systems?,'' and ``what is the fundamental nature of the gravitational interaction?''~\cite{LIGOScientific:2016lio, LIGOScientific:2018dkp, LIGOScientific:2019fpa, LIGOScientific:2020tif, LIGOScientific:2021sio}. Most of the current detections are consistent with the objects being black holes~(BHs) as described by general relativity~(GR)~\cite{LIGOScientific:2021sio}, which obey the no-hair theorems~\cite{Hawking:1972,Israel:1967,Israel:1968,Carter:1971,Robinson:1975}, implying that the multipolar structure of BHs in GR is highly constrained. This is not the case if nature is not described by GR, or if the binary components are not BHs (e.g. deformed stars~\cite{Raposo:2020yjy} or exotic compact objects~\cite{Cardoso:2019rvt}). In such cases, the GW signal can deviate from the standard binary BH~(BBH) vacuum-GR scenario, and the signal emitted during the inspiral can encode the intricacies of the objects' structure, allowing for measurements of their properties.

In the inspiral phase of spin-aligned quasi-circular binaries, these deviations typically enter at integer powers of the orbital velocity in both the GW phase and amplitude, corresponding to specific post-Newtonian~(PN) orders~\cite{Tahura:2018zuq}. A theory agnostic approach to testing GR with GWs is then to include a single deviation at a given PN order, with an arbitrary parameter that maps to specific theories. This approach led to the development of the \textit{parameterized post-Einsteinian framework}~(ppE)~\cite{Yunes:2009ke}, which has been a powerful tool for mapping theory agnostic constraints into constraints on beyond-GR effects~\cite{Yunes:2016jcc, LIGOScientific:2016lio, LIGOScientific:2021sio}.

As powerful as such framework is, degeneracies among the physical parameters of the binary can prevent stringent constraints on the coupling constants of some modified theories of gravity~\cite{Nair:2019,Perkins:2021mhb}. Dynamical Chern-Simons~(dCS) gravity~\cite{Jackiw:2003pm,Alexander:2009tp}, a parity violating theory of gravity, is one such theory. Due to its parity violating nature, modifications to GR are intricately related to deviations from spherical symmetry, and hence, to the spin angular momentum of compact objects. Without accurate measurements of the spins in BBHs, one cannot place pure GW constraints on the coupling constant of the theory; an approach that combined the results of GW170817~\cite{LIGOScientific:2017vwq} and observations of the pulsar PSR J0030+0451~\cite{Lommen_2000,Arzoumanian_2018} was able to place the first constraints on dCS gravity as an effective field theory~\cite{Silva:2020acr}. It has been proposed that dCS gravity can be tested purely with GWs if one observes spin-precessing binaries~\cite{Alexander:2017jmt}, which are known to (at least partially) break parameter degeneracies~\cite{Chatziioannou:2014coa}. 

Precession dynamics are not only a result of misalignment between angular momenta, but can also be induced by compact objects with rich multipolar structure of their exterior spacetime geometry~\cite{PoissonWill,Loutrel:2022ant}. BHs in GR are expected to have a unique multipolar structure, where all $l\ge2$ multipoles only depend on the BH's mass and spin~\cite{Heusler-lrr,Geroch:1970,Hansen:1974,Carter:1971,Hawking:1972}. However, other compact objects (in particular exotic compact objects) can have a richer multipolar structure~\cite{Raposo:2018xkf}, including axisymmetry breaking~\cite{Herdeiro:2020kvf,Bena:2020see,Bena:2020uup,Bianchi:2020bxa,Bianchi:2020miz,Bah:2021jno}. To leading PN order, nonaxisymmetry is encoded in the $m=\pm1$ and the $m\pm2$ harmonics of the mass quadrupole, which are identically zero for Kerr BHs. Thus, measurements of nonaxisymmetry in a compact object's multipole structure would provide a smoking gun for beyond-vacuum/beyond-GR\footnote{Such a smoking gun does not necessarily imply that nature is not described by GR, although this is a possibility. It could also imply that the compact objects are comprised of some exotic field, but still be solutions to the Einstein field equations, and this is why we make this clarification. For brevity, hereafter, we will simply refer to such scenarios as just beyond-GR.} effects.

Generically, precession of the orbital plane induces an amplitude and phase modulation in the GWs emitted by the binary, because GWs are emitted preferentially along the orbital angular momentum~\cite{Apostolatos:1994}. The study of the effects of precession on GWs has a rich history, but in the context of spin-precessing BBHs, analytic Fourier domain waveforms were developed in~\cite{Chatziioannou:2017tdw} and extended to full inspiral-merger-ringdown~(IMR) waveforms in~\cite{Khan:2020,Khan:2019kot,Khan:2019}. With such a promising development, it is simple to ask whether such waveforms can be parameterized in such a way as to allow for probes of fundamental physics, similar to what was done for spin-aligned waveforms with the ppE formalism. Indeed, the answer is yes, and recently, precessing waveforms were developed in two beyond-GR scenarios that have already been described here, namely dCS gravity~\cite{Loutrel:2022tbk} and generic mass quadrupole effects~\cite{Loutrel:2022ant}.

Using these two scenarios as motivation, we here develop an extension of the original ppE framework to include the effects of precession. The original ppE framework was parameterized by four beyond-GR parameters, two exponent parameters that determine the PN order of the deviations and two amplitude parameters that determine the ``magnitude" of the deviations, with one set entering the GW phase and another the amplitude. The ppE phase parameters determine the type of GR modification that is being considered. Therefore, once they are fixed, the ppE amplitude parameters can be added to the other parameters of the binary and searched over in a parameter estimation or a model selection study. The strength of the ppE formalism is that once a constraint on the ppE amplitude parameters is obtained, one can directly map these bounds to constraints on the coupling constants of a very large set of specific modified theories of gravity~\cite{Yunes:2009ke,Chatziioannou:2012rf,Tahura:2018zuq,Mezzasoma:2022pjb}.

The precessing ppE framework developed herein requires more parameters to achieve the desired outcome. Unlike spin-aligned waveforms, the amplitude of precessing waveforms are time- (or frequency-) dependent, and care must be taken when handling their deviations. With the definitions of Wigner-D matrices and application of Fourier decompositions, the amplitudes can be split into a \textit{precession phase} that varies on the precession timescale, and a $\textit{spectral amplitude}$ that only varies on the radiation-reaction timescale. Generally, the precession phase and spectral amplitudes, and their beyond-GR extensions, are complicated functions of the frequency that can change depending on what effects are creating the precession. To simplify this, we PN expand the beyond-GR part of both of these quantities, because they are already suppressed by non-GR coupling parameters that are expected to be small.

After PN expanding the precession quantities, we are still left with waveforms that contain beyond-GR effects in all harmonics, in contrast to the simplicity of the original ppE formalism. How does one determine which harmonics should be parameterized by ppE parameters? To determine this, we compute the likelihood between precessing waveforms in GR, and waveforms containing ppE deformations in one harmonic at a time. This allows us to determine which harmonic produces the largest contribution to the likelihood, and to determine how many and which harmonics are needed to achieve a desired percentage of the total likelihood (which we take to be 90\%). From this analysis, we determine that dCS waveforms need ppE parameters in three harmonics, while for non-axisymmetric quadrupole, ppE parameters in four harmonics are necessary. 

This paper, therefore, serves as a basis for the construction of a spin-precessing ppE model. With this model in hand, one could envision, for example, the inclusion of these ppE deformations to the {\tt IMRPhenompv3} waveform model~\cite{Khan:2020,Khan:2019kot,Khan:2019} and the study of GR tests with LIGO/Virgo data. Further work will be required to determine the minimum number of harmonics that are truly necessary to carry out conservative tests of GR in practice. Degeneracies between parameters can lead to a non-trivial structure of the likelihood function, which will necessitate a Bayesian study. We leave this, and much more, to future work. 

The main results of the paper are as follows: the full precessing ppE waveform is given in Eqs.~\eqref{eq:prec-ppE}, the spectral amplitudes are given in ppE form in Eqs.~\eqref{eq:PK-expand}-\eqref{eq:amp-ppE}, and the ppE waveform phase is given in Eq.~\eqref{eq:ppE-phase}-\eqref{eq:ppE-phase-2}. In Sec.~\ref{sec:rev}, we provide a general overview of precessing waveforms and the two example scenarios that motivate this study. In Sec.~\ref{sec:ppE}, we develop the precessing ppE waveforms and show how the parameters map to the two example scenarios. Finally, in Sec.~\ref{sec:disc}, we discuss the future prospects of this work, particularly in regards to parameter estimation and tests of GR. Throughout this paper, we use the convention $G=1=c$.

\section{\label{sec:rev}Beyond-GR Precessing Waveforms}

Before considering the development of a ppE formalism for precessing binaries, we provide a brief review of the currently known beyond-vacuum/GR precessing waveforms (see Refs.~\cite{Loutrel:2022ant,Loutrel:2022tbk} for more details). We begin with a broad description of precessing waveform models that describe the inspiral of compact binaries, with an emphasis on the formalism applicable to {\tt IMRPhenomPv3}. We then specialize our discussion to precessing waveform models beyond GR, first focusing on dCS gravity and then on compact objects with non-axisymmetric quadrupole deformations.

\subsection{The Anatomy of a Precessing Waveform}
All precessing waveforms of the inspiral phase of the coalescence in the PN approximation can be written as
\begin{align}
    \label{eq:h-def}
    \tilde{h}(f) = \sqrt{\frac{2}{3}} \frac{{\cal{M}}^{5/6}}{D_{L} \pi^{1/6}} f^{-7/6} \sum_{l=2}^{\infty} \sum_{m < 0 } {\cal{A}}_{lm}(f) e^{i\tilde{\Psi}_{m}(f)}
\end{align}
where ${\cal{M}}$ is the chirp mass of the binary, $D_{L}$ is the luminosity distance from the detector to the source, and $f$ is the Fourier frequency\footnote{Here, the sum only extends over negative $m$ due to the stationary phase condition. The phase of the Fourier integral is $\Psi_{f} = 2\pi f t(u) + m \phi(u)$, with $u=(2\pi M F)^{1/3}$. The stationary phase condition $d\Psi_{f}/dt$ is satisfied for positive frequencies when $m<0$ and negative frequencies when $m>0$. Note that waveforms computed in the time domain and numerically Fourier transformed will have all $m$ harmonics.}. The function $\tilde{h} = \tilde{h}_{+} - i \tilde{h}_{\times}$ is the Fourier domain waveform, with $h_{+,\times}$ the plus and cross GW polarizations.

The Fourier phase of the waveform is computed by the radiation reaction equation describing the evolution of the orbital frequency, or more specifically the PN expansion variable $u = (2\pi M F)^{1/3}$, with $F$ the orbital frequency. The PN coefficients of $du/dt$ can depend on the precession timescale, making them oscillatory and complicating the analysis to obtain $u(t)$. To solve this problem, a multiple scale analysis can be employed since there is a separation between the precession and radiation reaction timescales during the inspiral. The oscillatory effects are characterized by the nutation phase $\psi$, which is purely secular on the radiation reaction timescale. The leading-order terms in the multiple scale analysis~(MSA) are then given by the precession average of $du/dt$ over $\psi$, which can readily be computed if one knows the analytic solution to the precession dynamics under consideration. The phase of the Fourier integral is then $\Psi = 2\pi f t(u) + m \phi(u)$, with $t$ the time variable and $\phi$ the orbital phase. The integral can be computed by combining the stationary phase approximation~(SPA) and the shifted uniform asymptotic~(SUA) method of~\cite{Klein:2014bua,Chatziioannou:2017tdw}, which gives the Fourier phase
\begin{align}
    \label{eq:spa-def}
    \tilde{\Psi}_{m}(f) &= 2\pi f t_{c} + m \phi_{c} - \frac{\pi}{4} 
    \nn \\
    &- \frac{3m}{256\eta \tilde{u}^{5}} \left\{1 + \sum_{n}^{\infty} \left[\langle \Psi_{n}\rangle_{\psi} + \langle \Psi_{n}^{l}\rangle_{\psi} \ln \tilde{u}\right] \tilde{u}^{n}\right\}
\end{align}
where $\tilde{u} = (2\pi M f/|m|)^{1/3}$, $(t_{c},\phi_{c})$ are the time and phase of coalescence, $\eta$ is the symmetric mass ratio of the binary, and the coefficients $[\langle \Psi_{n}\rangle_{\psi}, \langle \Psi_{n}^{l}\rangle_{\psi}]$ are known functions of the binary's parameters and beyond-vacuum/GR parameters.

\begin{figure}[hbt!]
    \centering
    \includegraphics[width=\columnwidth, trim = 6cm 4cm 6cm 2cm, clip]{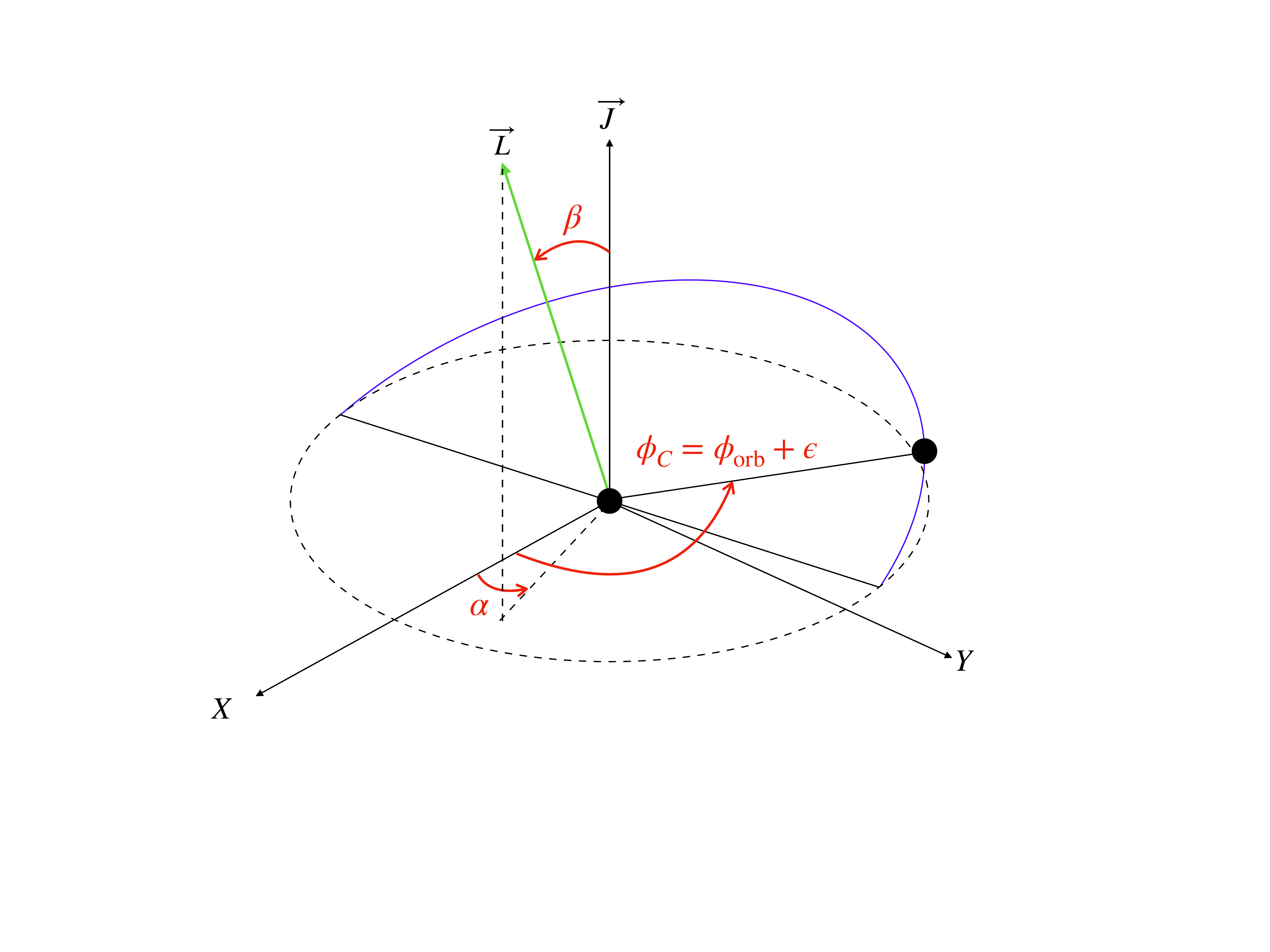}
    \caption{Geometric setup of the binary system considered herein. $\vec{J}$ is the total angular momentum of the binary and is fixed along the Z-axis. $\vec{L}$ is the orbital angular momentum, and is oriented by the inclination angle $\beta$ from $\vec{J}$, and the precession angle $\alpha$ from the X-axis. The carrier phase of the binary is the angle $\phi_{C}$ between the X-axis and the radial separatrix of the two BHs in the orbital plane, and is the sum of the orbital phase $\phi_{\rm orb}$ and Thomas phase $\epsilon$.}
    \label{fig:prec}
\end{figure}

For the waveform amplitudes, the particular combination $h_{+} - ih_{\times}$ allows for a decomposition of the time domain amplitudes $H_{lm}(u)$ into Wigner-D matrices ${D^{l}}_{mm'}$ (discussed later on), specifically
\begin{equation}
    H_{lm}(u) = \sum_{m'=-l}^{l} (-1)^{m'+1} {D^{l}}_{mm'}\left(\epsilon, \beta, -\alpha\right) {_{-2}}Y_{lm'}\left(\theta_{N},\phi_{N}\right)\,.
\end{equation}
where $\epsilon$ is the Thomas phase, $\beta$ is the inclination angle between the orbital angular momentum $\vec{L}$ and the total angular momentum $\vec{J}$, $\alpha$ is the precession angle with respect to the fixed axis defined by $\vec{J}$, and $(\theta_{N},\phi_{N})$ define the orientation of the line of sight vector with respect to $\vec{J}$. We parameterize the amplitude in terms of $[\alpha,\beta,\epsilon]$ since these are the precession phases appearing in the {\tt IMRPhenomPv3}~\cite{Khan:2020,Khan:2019kot,Khan:2019} waveforms. Figure~\ref{fig:prec} displays the geometric setup of this parameterization. 

The Fourier domain amplitudes are computed by the SUA method, specifically
\begin{align}
    \label{eq:amp-sua}
    {\cal{A}}_{lm}(f) &= \sum_{k=0}^{k_{\rm max}} \frac{a_{k,k_{\rm max}}}{2} \left[H_{lm}(u_{k}) + H_{lm}(u_{-k})\right]
\end{align}
where the coefficients $a_{k,k_{\rm max}}$ are defined by solving
\begin{equation}
\frac{(-i)^{p}}{2^{p} p!} = \sum_{k=0}^{\rm k_{\rm max}} a_{k,k_{\rm max}} \frac{k^{2p}}{(2p)!}\,,
\end{equation} 
for $p \in [0, k_{\rm max}]$, and $k_{\rm max}$ is chosen based on considerations of waveform accuracy and computational efficiency (see~\cite{Chatziioannou:2017tdw} for a detailed discussion). The SUA frequency function $u_{k}$ is defined as $u_{k} = u(t_{\star} + k T_{m})$, where $t_{\star}$ is the stationary point, and $T_{m} = [|m|\ddot{\phi}(t_{\star})]^{-1/2}$. Much like the Fourier phase, the functions $u_{k}$ can generically be written in a PN expansion of the variable $\tilde{u}$ as
\begin{align}
    \label{eq:sua-eq}
    u_{k} &= \tilde{u} + 4k \sqrt{\frac{2\eta}{15|m|}} \tilde{u}^{7/2} \left\{1 + \sum_{n} \left[\upsilon_{n} + \upsilon_{n}^{l} \ln \tilde{u}\right] \tilde{u}^{n}\right\}
\end{align}
where, once again, the coefficients $[\upsilon_{n}, \upsilon_{n}^{l}]$ depend on the binary's parameters and the beyond-vacuum/GR parameters. 

While the discussion of the waveform up to this point is complete, it is worth discussing the behavior of the Wigner-D matrices in more detail. From~\cite{Breuer}, the Wigner-D matrices can be decomposed into spin-weighted associated Legendre polynomials~(SALPs) as
\begin{align}
    {D^{l}}_{mm'}(\epsilon,\beta,-\alpha) &= (-1)^{m'} N_{lm} e^{im\epsilon + i m'\alpha} {_{m'}}P_{lm}(\cos\beta)\,,
\end{align}
where
\begin{align}
    N_{lm} &= \sqrt{\frac{(l-m)!}{(l+m)!}}\,.
\end{align}
The inclination angle $\beta$ is generically oscillatory in the nutation phase $\psi$, and it is thus convenient to decompose $\beta$, or functions of $\beta$, in a Fourier series. As a result, the SALPs can be decomposed as
\begin{align}
    \label{eq:amp-fourier}
    {_{m'}}P_{lm}(\cos\beta) &= \sum_{n=-\infty}^{\infty} {P^{l}}_{mm'n}(u) e^{in\psi}
\end{align}
where the ${P^{l}}_{mm'n}$ are only functions of the PN expansion variable $u$. Thus, the amplitudes $H_{lm}(u)$ can be written as
\begin{align}
    \label{eq:amp-H}
    H_{lm}(u) &= - N_{lm} \sum_{m'n} {P^{l}}_{mm'n}(u) e^{i\Phi^{\rm P}_{lmm'n}(u)} \,,
\end{align}
where
\begin{align}
    \label{eq:prec-phase}
    \Phi^{\rm P}_{lmm'n}(u) = m \epsilon(u) + m' \alpha(u) + n \psi(u)\,,
\end{align}
with the Fourier amplitude given by Eq.~\eqref{eq:amp-sua}.  As a matter of simplicity, we suppress the full harmonic index on these quantities by defining the hyper-index $K$ such that ${P^{l}}_{mm'n} = P_{K}$ and $\Phi^{\rm P}_{lmm'n} = \Phi^{\rm P}_{K}$.

Below, we review the details of the waveforms in two different contexts: dCS gravity, a parity violating modified theory of gravity, and generic quadrupole effects within GR. 

\subsection{Precessing waveforms in dCS gravity}
\label{sec:dcs}

The simplest theory of gravity with parity violating interactions that passes Solar System and binary pulsar tests is dCS gravity~\cite{Jackiw:2003pm, Alexander:2009tp}, which couples a psuedo-scalar field $\vartheta$ to the parity odd quadratic curvature invariant known as the Pontryagin density. Kerr BHs are modified in this theory, developing a scalar dipole moment that is proportional to the BH's spin, and modifies the higher order multipoles of the BH, thus violating the no-hair theorems~\cite{Yunes:2009hc,Yagi:2012ya,Maselli:2017kic}. The dynamics of BH binaries are modified to leading PN order through a modification to the Newtonian quadrupole-monopole interaction, a new conservative dipole-dipole interaction, and the emission of dipole radiation, the latter two not being present in GR. As a result of this, the spin precession equations for BBHs in dCS gravity were derived in~\cite{Loutrel:2018ydv}, with analytic waveforms for precessing binaries recently developed in~\cite{Loutrel:2022tbk}. 

The analytic solution of the spin precession problem in dCS gravity is characterized by an effective field theory extension of the solutions in GR, which were found in~\cite{Chatziioannou:2017tdw}. This is characterized by a weak coupling expansion, where one defined a dimensionless coupling parameter, and performs perturbation theory about the GR solution with this parameter being small. The theory is described by the dimensionful coupling parameter $\xi$, and for astrophysical systems, the dimensionless coupling parameter is $\zeta=\xi/M^{4}$, where $M$ is the mass scale of the system under consideration. In~\cite{Loutrel:2022tbk}, this was defined as $\zeta_{2} = \xi/m_{2}^{4}$, where $m_{2}$ is the mass of the smaller BH. Note that this effective field theory treatment of dCS gravity is required, due to issues related to the presence of higher-order derivatives in the field equations and well-posedness of the initial value problem in this theory~\cite{Delsate:2014hba}. However, when considering the spin precession problem in this theory, a non-uniform expansion~\cite{Bender} arises when considering the dCS coupling parameter $\zeta_{2}$ to be small, and simultaneously taking the equal-mass limit. This results from the fact that the no-hair theorems in GR imply that nutation is not present in equal-mass precessing binaries. However, this is not true in dCS gravity, and thus nonuniformity in the weak coupling approximation arises. The solution to this is to work in terms of an effective coupling parameter $\bar{\zeta}_{2} = \zeta_{2}/(1-q)^{2}$, and perform perturbation theory by requiring that $\bar{\zeta}_{2} \ll 1$.

The first step of the computation is to solve for the evolution of the total spin magnitude of the binary $S^{2}$ in a co-precessing frame. The evolution of this quantity physically encodes nutation into the GWs, and is oscillatory with a phase variable $\psi$ that evolves on the radiation reaction timescale. Through an MSA, this nutation phase takes the form
\begin{align}
    \label{eq:psi-dcs}
    \psi &= \psi_{c} - \frac{5}{128} \frac{(1-q^{2})}{q} u^{-3} 
    \nn \\
    &\times \left[1 + \sum_{n} \left(\psi_{n} + \psi_{n}^{l} \ln u\right) u^{n} + \bar{\zeta}_{2} \delta \psi_{2} u^{2}\right]
\end{align}
where the PN coefficients $(\psi_{n}, \psi_{n}^{l})$ are given to 1PN order in Eqs.~\eqref{eq:psi1}-\eqref{eq:psi2}, and more generally in Appendix~D of~\cite{Loutrel:2022tbk}. The coefficient of the dCS correction, i.e. the term proportional to $\bar{\zeta}_{2}$ above, is given in Eq.~\eqref{eq:dpsi2}, and enters at relative 1PN order. This may seem strange since the spin precession equations in Eq.~\eqref{eq:prec} are actually modified at relative 0.5PN order, corresponding to correction to the spin-spin and quadrupole-monopole couplings. The reason for this is that the general solution for $S^{2}$ is given in terms of Jacobi Elliptic functions $\text{sn}(x,y)$, specifically
\begin{align}
    \label{eq:Ssq}
    S^{2} = S_{+}^{2} + (S_{-}^{2} - S_{+}^{2})\text{sn}^{2}(\psi, \cm)\,,
\end{align}
where $(S_{-}^{2},S_{+}^{2})$ are the minimum and maximum values of $S^{2}$, and $\cm$ is the modulus of the oscillations. As their name might suggest, Jacobi elliptic functions have oscillations that are elliptic in nature, as opposed to circular for standard trigonometric functions, and this ellipticity is dependent on the modulus $\cm$. In the case of spin precession, $\cm \sim u^{2}$, which is modified at relative Newtonian order in dCS gravity. Due to its scaling with $u$, the modulus $\cm$ modifies the nutation phase at 1PN order, and thus the dCS correction appears at relative 1PN order. As a final note, unlike the other angles in the spin precession problem, $\psi$ is a purely secular function on the radiation reaction timescale, i.e. it is non-oscillatory.

The inclination angle $\beta$ for this scenario is directly related to the angular momenta $(J, L, S^{2})$ through
\begin{align}
    \label{eq:cosbeta}
    \cos\beta = \frac{J^{2} + L^{2} - S^{2}}{2J L}
\end{align}
where $S^{2}$ is given by Eq.~\eqref{eq:Ssq}. Under radiation reaction, $(J, L, S_{\pm}^{2},\cm)$ all evolve in time, and can be written as functions of the PN expansion variable $u$, specifically
\begin{align}
    \label{eq:spm-pn}
    S_{\pm}^{2} &= s_{\pm}^{(0)} + \bar{\zeta}_{2} \delta s_{\pm}^{(0)} + {\cal{O}}(u)\,,
    \\
    \mathscr{m} &= \left(\frac{s_{-}^{(0)} - s_{+}^{(0)}}{s_{3}^{(0)}}\right) u^{2} \left[1 + \bar{\zeta}_{2}\delta \mathscr{m}+ {\cal{O}}(u)\right]
\end{align}
where the GR coefficients $s_{\pm,3}^{(n)}$ are given in Appendix B of~\cite{Loutrel:2022tbk}, and the dCS corrections $(\delta s_{\pm}^{(0)}, \delta\mathscr{m})$ are given in Eqs.~(54) \&~(57) therein, respectively. The mapping between $L$ and $u$ is given by the conservative dynamics of the binary, and in~\cite{Loutrel:2022tbk,Chatziioannou:2017tdw}, this was taken to be the Newtonian mapping $L = \eta M^{2}/u$. The evolution of $J$ under radiation reaction must be handled with an MSA, with the leading-order terms given by a precession average of the radiation reaction effects. The method to carry this out is explained in Sec.~IVA of~\cite{Loutrel:2018rxs}, with the result being
\begin{align}
    \label{eq:J-dcs}
    J^{2} = \sum_{n=0} j_{n} u^{n-2} + \frac{\bar{\zeta}_{2}}{2} \left(\delta s_{+}^{(0)} + \delta s_{-}^{(0)}\right)
\end{align}
where the $j_{n}$ coefficients are
\begin{align}
    \label{eq:j-coeffs}
    j_{0} = M&^{4} \eta^{2}\,, \qquad j_{1} = 2c_{1} M^{2} \eta\,, 
    \nn \\
    j_{2} &= \frac{1}{2} \left(s_{+}^{(0)} + s_{-}^{(0)}\right)\,.
\end{align}
With this, Eq.~\eqref{eq:cosbeta} can be written in a Fourier decomposition of the purely secular nutation phase as
\begin{align}
    \label{eq:beta-dcs}
    \cos \beta = \sum_{k=-2}^{2} {\cal{C}}_{k} e^{ik\psi}
\end{align}
where the ${\cal{C}}_{k}$ are generically functions of $u$, given by
\begin{align}
    {\cal{C}}_{0} &= \frac{2J^{2} + 2L^{2} - S_{+}^{2} - S_{-}^{2}}{2JL}\,,
    \\
    {\cal{C}}_{\pm2} &= \frac{S_{-}^{2} - S_{+}^{2}}{8 J L}\,.
\end{align}
with $S_{\pm}^{2}$ given by Eq.~\eqref{eq:spm-pn} and $J$ is given by Eq.~\eqref{eq:J-dcs}. More general trigonometric functions of $\beta$ have more complicated structure, but can be written generically in terms of Gegenbauer polynomials. A full explanation of this can be found in Appendix F of~\cite{Loutrel:2022tbk}.

Once, the nutation phase $\psi$ is computed, the precession angle $\alpha$ can be found by rotating to a nonprecessing frame and forcing the evolution of $\alpha$ to obey the spin precession equations via the method described after Eq.~(34) in~\cite{Loutrel:2022tbk}. In general, the evolution of $\alpha$ is oscillatory on the precession timescale, but also evolves on the longer radiation reaction timescale. The evolution of $\alpha$ can then be solved through MSA, with the solution taking the form,
\begin{align}
    \alpha &= \alpha_{-1}(u) + \lambda \; \alpha_{0}(\psi, u) + {\cal{O}}(\lambda^{2})\,,
\end{align}
where $\alpha_{-1}$ is purely secular and only depends on $u$, $\alpha_{0}$ contains both secular and oscillatory contributions, and $\lambda \sim T_{\rm prec}/T_{\rm rr}$ is an order keeping parameter that scales as the ratio of the precession timescale to the radiation reaction timescale. There is one caveat to this analysis, however. The typical PN expansion ($u \ll 1$) of $d\alpha/dt$ results in a loss of accuracy for smaller mass ratios~\cite{Chatziioannou:2017tdw}. Instead, the calculation of $\alpha$ is performed by factoring out an overall prefactor of $J$ from $d\alpha/dt$, PN expanding the remainder, and then integrating over the radiation reaction timescale. The end result of this is a function of the form
\begin{align}
    \label{eq:alpha-dcs}
    \alpha_{-1}(u) &= \sum_{n=-3} \alpha_{n}^{(-1)} \varphi_{n}(u) 
    \nn \\
    &+ \bar{\zeta}_{2} \left[\delta \alpha_{-3}^{(-1)} \varphi_{-3}(u) + \alpha_{-3}^{(-1)} \delta \varphi_{-3}(u)\right]
\end{align}
where the GR coefficients $\alpha_{n}^{(-1)}$ are given by Eqs.~\eqref{eq:Phiz-3}-\eqref{eq:Phiz-2} and the functions $\varphi_{n}(u)$ are given by
\begin{align}
    \label{eq:phin}
    \varphi_{n} &= \int du \; J_{0}(u) \; u^{n}\,.
\end{align}
where $J_{0}(u) = \lim_{\bar{\zeta}_{2} \rightarrow 0} J(u)$, with $J(u)$ given in Eq.~\eqref{eq:J-dcs}. The dCS correction is then characterized by the coefficient $\delta \alpha_{-3}^{(-1)}$, which is given in Eq.~\eqref{eq:dPhiz-3}, and the function $\delta \varphi_{-3}(u)$ is given by
\begin{equation}
    \label{eq:dphi-3}
    \delta \varphi_{-3} = \int \frac{du}{J_{0}(u)} u^{-3}\,.
\end{equation}
A similar MSA calculation provides the evolution of the Thomas phase, specifically
\begin{align}
    \label{eq:eps-dcs}
    \epsilon_{-1}(u) &= \sum_{n=-3} \epsilon_{n}^{(-1)} u^{n} + \bar{\zeta}_{2} \delta \epsilon_{-3}^{(-1)} u^{-3}
\end{align}
where the coefficients of the GR sequence $\epsilon_{n}^{(-1)}$ are given in Eqs.~\eqref{eq:PhiT-0}-\eqref{eq:PhiT-1}, and the coefficient of the dCS correction $\delta \epsilon_{-3}^{(-1)}$ is given in Eq.~\eqref{eq:dPhiT-0}.

The SPA phase and SUA correction take the forms of Eqs.~\eqref{eq:spa-def} and~\eqref{eq:sua-eq}, respectively. However, unlike the case of nonprecessing quasi-circular binaries where beyond-GR corrections are captured by one correction, the precessing case requires three corrections for dCS gravity. The reason for this is that the GR 1.5PN spin-orbit and 2PN spin-spin couplings in the GW fluxes are modified in dCS gravity since they depend on the relative orientation of the spins and orbital angular momentum, which evolve differently in dCS gravity compared to GR. Yet, in the nonprecessing, spin-aligned (or anti-aligned) scenario, the dCS correction enters at 2PN order, resulting from the emission of dipole radiation and the near zone dipole-dipole force, the latter of which modifies the binding energy of the orbit~\cite{Yagi:2012vf}. Thus, the dCS corrections to the spin-orbit and spin-spin couplings have to be suppressed in that limit. This was indeed shown to be the case in~\cite{Loutrel:2022tbk}, so for a complete parameterization of the corrections, all three deviations were included. For the SPA phase in Eq.~\eqref{eq:spa-def}, the corrections are
\begin{align}
    \label{eq:spa-dcs-1}
    \langle \delta \Psi_{3} \rangle_{\psi}^{\rm dCS} &= 4\bar{\zeta}_{2} \langle \delta \beta_{3}^{\rm dCS} \rangle_{\psi}\,,
    \\
    \label{eq:spa-dcs-2}
    \langle \delta \Psi_{4} \rangle_{\psi}^{\rm dCS} &= \bar{\zeta}_{2} \left[10 \langle \delta \sigma_{4}^{\rm dCS} \rangle_{\psi} - 160 \frac{q(1-q)^{2}}{(1+q)^{4}} \langle \delta C\rangle_{\psi}\right]\,,
\end{align}
while the corrections to the SUA modification are
\begin{align}
    \delta \upsilon_{3}^{\rm dCS} &= -\frac{\bar{\zeta}_{2}}{2} \langle \delta \beta_{3}^{\rm dCS} \rangle_{\psi}\,,
    \\
    \delta \upsilon_{4}^{\rm dCS} &= \bar{\zeta}_{2} \left[-\frac{1}{2} \langle \delta \sigma_{4}^{\rm dCS} \rangle_{\psi} + 8 \frac{q(1-q)^{2}}{(1+q)^{4}} \langle \delta C\rangle_{\psi}\right]\,.
\end{align}
In the above expression, $\langle \delta \beta_{3}\rangle_{\psi}$ is the deviation to the spin-orbit coupling, $\langle \delta \sigma_{4} \rangle_{\psi}$ is the deviation to the spin-spin coupling, $\langle \delta C\rangle_{\psi}$ is the dCS dipole radiation term, and these are given in Eqs.~\eqref{eq:db3},\eqref{eq:ds4},\eqref{eq:dC-avg}, respectively.

The last piece needed to construct the precessing waveforms in dCS gravity are the Fourier amplitudes of the SALPs, specifically $P_{K}(u)$. In dCS gravity, these are complicated functions of the PN expansion variable due to their dependence on $J$ through Eqs.~\eqref{eq:cosbeta} and~\eqref{eq:J-dcs}, and do not follow a power series expansion. In general, the functions can be specified using the Gegenbauer polynomial decomposition described in Appendix~F of~\cite{Loutrel:2022tbk}, with the GR terms and dCS corrections given by
\begin{align}
    P_{K}^{\rm GR}(u) &= \lim_{\bar{\zeta}_{2} \rightarrow 0} P_{K}(u)\,,
    \\
    \label{eq:dPK-dcs}
    \delta P_{K}^{\rm dCS}(u) &= \lim_{\bar{\zeta}_{2} \rightarrow 0} \frac{\partial P_{K}(u)}{\partial \bar{\zeta}_{2}}\,.
\end{align}
The precessing waveforms are now fully specified in dCS gravity.

\subsection{Precessing waveforms for BHs with non-axisymmetric quadrupole deformations}
\label{sec:quad}

Reference~\cite{Loutrel:2022ant} considered the effect of generic mass quadrupole moments on the dynamics of a binary system. The mass quadrupole moments of generic compact objects can be decomposed into spherical harmonics, with the $m=0$ modes describing the standard spheroidal scenario (the spin-induced quadrupole as an example), and the $m=\pm1$ and $m=\pm2$ describing axial and polar modes, respectively. The latter of these are described by modulus $\cq_{m}$ and argument $\ca_{m}$ parameters\footnote{In~\cite{Loutrel:2022ant}, these are labeled $\epsilon_{m}$ and $\alpha_{m}$, respectively. We have changed the notation here to avoid confusion with the {\tt IMRPhenomPv3} phase variables.}. Isolated BHs in GR obey the no-hair theorems, and thus only possess the $m=0$ components of the mass quadrupole. However, horizonless compact objects or BHs that are not described by GR, need not necessarily obey these theorems, and can thus have axial and polar components to the mass quadrupole.

By working to relative Newtonian order in orbital, quadrupole, and radiation reaction effects, analytic solutions to the precession equations were obtained in~\cite{Loutrel:2022ant} by making use of the method of osculating orbits and MSA. The precession dynamics can generically be solved in terms of an auxiliary phase variable $\psi_{2}$, given by Eq.~(60) in~\cite{Loutrel:2022ant}. The solutions are exact for $0 \le \cq_{2} < 1$ and perturbative in $\cq_{1} \ll 1$. In keeping with the spirit of the original ppE formalism, we here consider the background solution to be that of a spheroidal quadrupole moment, and consider the polar and axial deviations to this scenario as small. The mapping between the {\tt IMRPhenomPv3} phase variable and those used in~\cite{Loutrel:2022ant} are $\Omega \leftrightarrow \alpha+\pi/2$, and  $\iota \leftrightarrow \beta$. 

The auxiliary phase $\psi_{2}$ is the main quantity in the precession dynamics that evolves under radiation reaction. However, the function itself is not purely secular; rather, it is oscillatory in terms of a purely secular function that we will label $\psi$ in analogy to the nutation phase in Sec.~\ref{sec:dcs}. This function is given by Eq.~(128) in~\cite{Loutrel:2022ant}, which to Newtonian order is specifically
\begin{align}
    \psi(u) &= \frac{3\sqrt{5\pi}}{32} \frac{Q_{0} \sqrt{1-\cq_{2}^{2}}}{M^{2} \eta u} \cos\beta_{0}\,,
\end{align}
where $Q_{0}$ is the $m=0$ component of the mass quadrupole, and $\beta_{0}$ is an integration constant associated with the inclination angle $\beta$. While the solutions in~\cite{Loutrel:2022ant} are exact for $0 \le \cq_{2} < 1$, we here treat the nonaxisymmetric deformations of the mass quadrupole as being small.

Similar to how the total spin magnitude is dependent on elliptic functions of $\psi$ in dCS gravity, the auxiliary phase is dependent on $\psi$ through elliptic integrals and Jacobi elliptic functions of modulus (see Eq.~\eqref{eq:psi2-pol})
\begin{align}
    \cm = \frac{2\cq_{2} \tan^{2}\beta_{0}}{1-\cq_{2}}\,.
\end{align}
The solution in terms of elliptic functions is only valid when $\cm < 1$. However, $\tan\beta_{0}$ can be arbitrarily large when $\beta_{0} \rightarrow \pi/2$. This implies that there is a critical limit to the inclination angle, specifically
\begin{align}
    \beta_{0} < \tan^{-1}\left[\left(\frac{1-\cq_{2}}{2\cq_{2}}\right)^{1/2}\right]\,.
\end{align}
For the remainder of the analysis, we assume that this bound holds. Under this assumption, taking the limit $\cq_{2} \ll 1$ corresponds to $\cm \ll 1$. Applying this to Eq.~\eqref{eq:psi2-pol} herein and Eq.~(130) in~\cite{Loutrel:2022ant} gives
\begin{align}
    \psi_{2}(u) &= \tilde{\psi}(u) + \frac{1}{4} \cq_{2} \tan^{2}\beta_{0} \sin[2\tilde{\psi}(u)] 
    \nn \\
    &+ \cq_{1} \tan\beta_{0} \cos[\Delta + \tilde{\psi}(u)] + {\cal{O}}(\cq_{m}^{2})\,,
\end{align}
where $\Delta = \ca_{1} - \ca_{2}$, and $\tilde{\psi}(u)$ is the re-summed nutation phase
\begin{align}
    \label{eq:nut-quad}
    \tilde{\psi}(u) &= \tilde{\psi}_{c} - \frac{3\sqrt{5\pi}}{32} \frac{\chi_{Q}}{u} \Big[\cos\beta_{0} - \frac{1}{4} \cq_{2} \tan^{2}\beta_{0} 
    \nn \\
    &- \cq_{1} \tan\beta_{0} \sin\Delta + {\cal{O}}(\cq_{m}^{2}) \Big]
\end{align}
with $\chi_{Q} = Q_{0}/M^{3}\eta$ and $\tilde{\psi}_{c}$ an integration constant. The benefit of re-summing the above expressions in this manner is that now $\psi_{2}$ and all precession phases can be written purely in terms of the secularly evolving $\tilde{\psi}$.

The full expression for the inclination angle $\beta$ is given in Eq.~\eqref{eq:beta-full}. Much like the case of dCS gravity in Sec.~\ref{sec:dcs}, it is simpler to write out the expansion of a trigonometric function of $\beta$ rather than the inclination angle itself. When expanding in $\cq_{1,2} \ll 1$, we have
\begin{align}
    \sin \beta = \sum_{k=-2}^{2} {\cal{S}}_{k} e^{ik\tilde{\psi}} + {\cal{O}}(\cq_{m}^{2})\,,
\end{align}
which is analogous to Eq.~\eqref{eq:beta-dcs} for dCS gravity. The main difference is that here the Fourier coefficients ${\cal{S}}_{k}$ are not functions of $u$, but depend only on integration constants and the quadrupole parameters $\cq_{m}$ and $\ca_{m}$. To linear order in the moduli $\cq_{m}$, we have
\begin{align}
    {\cal{S}}_{0} &= \sin\beta_{0} - \frac{1}{2} \cq_{2} \sin\beta_{0} + \cq_{1} \cos\beta_{0} \sin\Delta\,,
    \\
    {\cal{S}}_{1} &= {\cal{S}}_{-1}^{\dagger} = -\frac{i}{2} \cq_{1} e^{-i\Delta}  \cos\beta_{0}\,,
    \\
    {\cal{S}}_{\pm2} &= \frac{1}{4} \cq_{2} \sin\beta_{0}\,,
\end{align}
where $\dagger$ corresponds to complex conjugation. 

While in this case the MSA is not needed in order to calculate the precession angle $\alpha$, the latter can still be split into secular and oscillatory pieces much like the case of dCS gravity. The full expression for $\alpha$ in this case is given in Eq.~\eqref{eq:alpha-full}. After expanding in $\cq_{1,2} \ll 1$, the precession angle can also be written in a Fourier series as
\begin{align}
    \label{eq:alpha-quad}
    \alpha = -\tilde{\psi} + \sum_{k=-2}^{2} A_{k} e^{ik\tilde{\psi}} + {\cal{O}}(\cq_{m}^{2})\,,
\end{align}
with
\begin{align}
    A_{0} &= -\frac{\pi}{2} - \ca_{2} + \cq_{1} \cos\beta_{0} \sin\Delta\,,
    \\
    A_{1} &= A_{-1}^{\dagger} = -\frac{i}{2} \cq_{1} e^{-i\Delta} \left(\cos\beta_{0} - i \tan\beta_{0}\right)\,,
    \\
    A_{2} &= A_{-2}^{\dagger} = \frac{i}{8} \cq_{2} \left(2 + \tan^{2}\beta_{0}\right)\,.
\end{align}
There is a subtle difference here when comparing to the case of dCS gravity in Sec.~\ref{sec:dcs}. The secular behavior of $\alpha$ is purely given in terms of the nutation phase, which obeys a typical PN expansion from Eq.~\eqref{eq:nut-quad}. In the case of dCS gravity, specifically Eq.~\eqref{eq:alpha-dcs}, the precession angle cannot be written as a PN expansion in power of $u$. Instead, it is written in terms of the functions given in Eq.~\eqref{eq:phin}. The reason for this is that the precession dynamics in the quadrupole scenario are significantly simpler and only consider the effect of generic mass quadrupoles. The spin precession problem in both GR and dCS gravity models all of the effects to order spin-squared in the precession equations, which causes the analysis to be more complicated from an analytic perspective.

In~\cite{Loutrel:2022ant}, the Thomas phase was not explicitly calculated, but we provide its analytic expression here in the limit $\cq_{m} \ll 1$. In the osculating formalism, the true anomaly $V$ of the orbit is analogous to the carrier phase $\phi_{C}$ for precessing systems, and obeys the equations $\dot{V} = \dot{\phi}_{\rm orb} + \dot{\omega} + \dot{\alpha}\cos\beta$, where $\phi_{\rm orb}$ is the orbital phase, $\omega$ is the longitude of pericenter, the overdot corresponds to differentiation with respect to time, and the latter two terms constitute the evolution of the Thomas phase, specifically $\dot{\epsilon} = \dot{\omega} + \dot{\alpha}\cos\beta$. The orbital phase evolves on the radiation reaction timescale and obeys Eq.~(124) in~\cite{Loutrel:2022ant}. The longitude of pericenter is explicitly given in terms of the auxiliary phase $\psi_{2}$ in Eq.~\eqref{eq:omega-full}. The contribution arising from $\dot{\alpha}\cos\beta$ can be readily computed from Eqs.~\eqref{eq:alpha-dot} \&~\eqref{eq:beta-full} by working in the limit $\cq_{m}\ll 1$. The end result is
\begin{align}
    \label{eq:eps-quad}
    \epsilon &= \Omega_{\epsilon} \tilde{\psi} + \sum_{k=-2}^{2} E_{k} e^{ik\tilde{\psi}} + {\cal{O}}(\cq_{m}^{2})
\end{align}
with
\allowdisplaybreaks[4]
\begin{align}
    \Omega_{\epsilon} &= \frac{1}{4} \sec\beta_{0} \left[1 + 3 \cos(2\beta_{0})\right] 
    \nn \\
    &-\frac{1}{4} \cq_{1} \sec\beta_{0} \tan\beta_{0} \left[5 + 3 \cos(2\beta_{0})\right]\sin\Delta 
    \nn \\
    &+ \frac{1}{32} \cq_{2} \sec^{3}\beta_{0} \left[9 + 20 \cos(2\beta_{0}) + 3\cos(4\beta_{0})\right]\,,
    \\
    E_{0} &= \epsilon_{0} + 2 \cq_{2} \sec\beta_{0}
    \nn \\
    &-\frac{1}{32} \cq_{1} \left[14 \cos\Delta + 5 \cos(\Delta - 4 \beta_{0}) + 4 \cos(\Delta - 2\beta_{0}) 
    \right.
    \nn \\
    &\left.
    + 4 \cos(\Delta+2\beta_{0}) + 5 \cos(\Delta+4\beta_{0})\right]\csc\beta_{0} \sec^{2}\beta_{0}
    \\
    E_{\pm 1} &= -\cq_{2} \sec\beta_{0}\,,
    \\
    E_{2} &= E_{-2}^{\dagger} = \frac{i}{4} \cq_{2} \sec\beta_{0}\,.
\end{align}
where $\epsilon_{0}$ is an integration constant. All of the precession angles are now fully specified.

In GR, the quadrupole-monopole interaction generically enters the GW phase at 2PN order. Thus, the corrections to the SPA phase in Eq.~\eqref{eq:spa-def} due to nonaxisymmetric quadrupole effects are of 2PN order, specifically
\begin{align}
    \label{eq:spa-q}
    \langle \delta \Psi_{4}\rangle_{\psi} &= \frac{5}{4} \sqrt{\frac{\pi}{5}} \chi_{Q} \left[\cq_{1} {\cal{U}}_{10} + \cq_{2} {\cal{U}}_{01} + {\cal{O}}(\cq_{m}^{2})\right]\,,
\end{align}
where ${\cal{U}}_{10}$ and ${\cal{U}}_{01}$ only depend on constants of integration and the modulus parameters $\ca_{m}$, and are given in Eqs.~(B27)-(B28) in~\cite{Loutrel:2022ant}. Similarly, the SUA correction to Eq.~\eqref{eq:sua-eq} is
\begin{align}
    \label{eq:sua-q}
    \delta \upsilon_{4} &= \frac{1}{16} \sqrt{\frac{\pi}{5}} \chi_{Q} \left[\cq_{1} {\cal{U}}_{10} + \cq_{2} {\cal{U}}_{01} + {\cal{O}}(\cq_{m}^{2})\right]\,.
\end{align}
Unlike the case of dCS gravity, the corrections to the phase are only captured by one correction. The reason for this is that the analysis carried out in~\cite{Loutrel:2022ant} did not explicitly include the spin contributions. However, we expect that once these are included, additional terms may be needed in the phase to capture all of the corrections, just like the scenario of dCS gravity. 

Lastly, the Fourier amplitudes of the SALPs in the nonaxisymmetric quadrupole scenario are
\begin{align}
    P_{K}^{\rm GR} &= \lim_{\cq_{m}\rightarrow0} P_{K}\,,
    \\
    \label{eq:dPK-q}
    \delta P_{K}^{\cq} &= \lim_{\cq_{1,2}\rightarrow0}\left(\frac{\partial P_{K}}{\partial \cq_{1}} + \frac{\partial P_{K}}{\partial \cq_{2}}\right)\,,
\end{align}
where it should be understood that the limits are taken with respect to both $\cq_{1}$ and $\cq_{2}$, and the corrections $\delta P_{K}^{\cq}$ are summed over both contributions. In this scenario, these Fourier amplitudes are not functions of the PN parameter $u$, as opposed to the scenario of dCS gravity. Here, they are only functions of the quadrupole parameters $(\cq_{m},\ca_{m})$ and constants of integration. This is also a result of the computations within~\cite{Loutrel:2022ant} not including spin effects, and we expect this to change once those effects are included. This completes our review of the precessing waveforms.

\section{\label{sec:ppE} Precessing ppE Framework}

In this section, we introduce the precessing ppE framework. We begin by presenting some basic considerations that one must take into account when constructing such a framework. We then detail the construction of the framework, separating the discussion of the Fourier phases to that of the Fourier amplitudes. We proceed by presenting explicit mappings of the ppE framework to the predictions of dCS gravity and that of compact objects with non-axisymmetric quadrupole moments. We conclude this section with a discussion of likelihoods and overlaps to decipher how many harmonics in the ppE deformations need to be kept.  

\subsection{Basic Considerations}
For non-precessing binaries, a large class of perturbative deviations from GR in both the phase and amplitude of the waveform can be written as PN corrections of the form~\cite{Yunes:2009ke}
\begin{equation}
    \label{eq:ppE}
    \tilde{h}(f) = \tilde{h}_{\rm GR}(f) \left[1 + \alpha_{\rm ppE} \left(\pi M f\right)^{a_{\rm ppE}}\right] e^{i \beta_{\rm ppE} \left(\pi M f\right)^{b_{\rm ppE}}}
\end{equation}
where $\tilde{h}_{\rm GR}(f)$ is the Fourier domain waveform in GR, $(\alpha_{\rm ppE}, \beta_{\rm ppE})$ are constants that depend on the parameters of the binary and the coupling constants of non-GR effects, and $(a_{\rm ppE}, b_{\rm ppE}) = (\tilde{a}_{\rm ppE}/3, \tilde{b}_{\rm ppE}/3)$ with $(\tilde{a}_{\rm ppE}, \tilde{b}_{\rm ppE})$ integers. Equation~\eqref{eq:ppE} constitutes the simplest incarnation of the ppE formalism in the inspiral regime, and can be mapped to a wide variety of modified theories of gravity~\cite{Yunes:2016jcc}. Therefore, the above waveform provides a theory agnostic model to perform tests of GR with GW observations, which can then be mapped to theory specific constraints once one knows the mapping between $(\alpha_{\rm ppE}, \beta_{\rm ppE})$ and the coupling constants of the theory under consideration. The usefulness of this formalism has already been amply shown with the current GW detections~\cite{Yunes:2016jcc,LIGOScientific:2016lio,LIGOScientific:2018dkp,LIGOScientific:2019fpa,LIGOScientific:2020tif,LIGOScientific:2021sio}, and will likely continue to be of relevance for future events~\cite{Perkins:2022fhr}. The question now is the following: can this formalism be extended to include precession?

Before we consider how to answer this question, it is important to lay out some of the most important and different modifications due to non-GR effects that appear when considering  precessing binaries. The list of non-GR modifications includes the following:
\begin{itemize}
    \item \textit{Near zone effects}: Modifying the action introduces new forces within the near zone of the binary, altering the orbital energy, orbital angular momentum, and Kepler's third law. When multiple effects are competing, for example between forces generated by new scalar/vector/tensor fields and modified multipole moments of the BHs, the lowest PN order corrections are those of most relevance.
    \item \textit{Precession effects}: In general, modifications to the precession equations alter their solution in the manner described here for dCS gravity, but their PN order may change in other theories. When calculating various phases, one typically has to employ an MSA and consider the precession average of all necessary quantities. During the latter, precession effects may back react onto those previously mentioned and force corrections at lower PN order than expected. This occurs in dCS gravity, as explained in Sec.~\ref{sec:dcs}, but it is not guaranteed to occur in all modified theories.~\footnote{As an example, suppose the 1.5PN spin-orbit coefficients $\beta_{3}$ is modified in Einstein-scalar-Gauss-Bonnet theory (another theory with a scalar field coupled to quadratic-curvature terms in the action) similar to how it is modified here for dCS gravity when taking the precession average. If so, the $du/dt$ in Einstein-scalar-Gauss-Bonnet theory would not be modified from what one expects in the non-precessing case, since the leading PN order correction appears at $-1$PN order due to dipole radiation.}
    \item \textit{SUA modifications}: The SUA resummation procedure shifts the SPA frequency of the waveform amplitudes by a correction that depends on $\ddot{\Phi}^{-1/2}$, where $\Phi$ is the phase of the time domain waveform. non-GR effects appear at standard PN orders relative to the leading PN order effect in GR. The relative PN order that non-GR modification appear in the SUA mapping are the same as those for $du/dt$, since $\Phi = \Phi(u)$.
\end{itemize}
These different effects all compete with one another, and care needs to be taken when mapping theory specific predictions to a precessing ppE framework. We will detail how to do so later in this section.

\subsection{Developing the Framework}

The primary difference between non-precessing and precessing waveforms in modified gravity is that the amplitudes are no longer monotonic functions that vary on the radiation reaction timescale, and thus can no longer be written in the format of Eq.~\eqref{eq:ppE}. However, the waveform amplitude can be Fourier decomposed into harmonics of $\psi$ through Eq.~\eqref{eq:amp-fourier}. As a result, the amplitude is split into monotonic functions $(P_{K}^{\rm GR}, \delta P_{K})$ that vary on the radiation-reaction timescale, and phase corrections $(\Phi^{\rm P, GR}_{K}, \delta \Phi^{\rm P}_{K})$ that vary on the precession timescale through Eq.~\eqref{eq:amp-H}. More specifically,
\begin{align}
    \label{eq:P-decomp}
    P_{K}^{\rm GR}(u) &= \lim_{\zeta^{A} \rightarrow 0} P_{K}(u)\,,
    \qquad
    \delta P_{K}(u) = \lim_{\zeta^{A} \rightarrow 0} \frac{\partial P_{K}(u)}{\partial \zeta^{A}}\,,
    \\
    \label{eq:Phi-decomp}
    \Phi^{\rm P,GR}_{K} &= \lim_{\zeta^{A} \rightarrow 0} \Phi^{\rm P}_{K}(u)\,,
    \qquad 
    \delta \Phi^{\rm P}_{K} =  \lim_{\zeta^{A} \rightarrow 0} \frac{\partial \Phi^{\rm P}_{K}(u)}{\partial \zeta^{A}}\,,
\end{align}
where $\Phi^{\rm P}_{K}(u)$ is given in Eq.~\eqref{eq:prec-phase} and $\zeta^{A}$ is shorthand for any (small) non-GR parameters ($\bar{\zeta}_{2}$ in dCS gravity, $\cq_{m}$ for non-axisymmetric quadrupoles). In the case of the considered scenario having multiple $\zeta^{A}$, one would sum over all corrections linear in $\zeta^{A}$.

\subsubsection{Phases}
\label{sec:phase-ppE}

In Sec.~\ref{sec:rev}, we explicitly showed that the SPA part of the Fourier phase produces the standard result one might expect from quasi-circular binaries, specifically
\begin{equation}
    \label{eq:spa-ppE}
    \tilde{\Psi}_{m}(f) = \tilde{\Psi}^{\rm GR}_{m}(f) + \mathscr{b} \tilde{u}^{b}\,,
\end{equation}
where $\Psi_{\rm GR}(f)$ is the GR sector of the phase, $
b$ is the exponent parameter and $\mathscr{b}$ is the associated amplitude parameter. For precessing binaries, the parameters $[\mathscr{b}, b]$ are not necessarily the same as those for spin-aligned binaries. For example, the lowest order correction to the SPA phase in dCS gravity for spin-aligned binaries is the 2PN dipole radiation correction, but for precessing binaries, it is the shift in the 1.5PN spin-orbit term. We will show this in more detail in Sec.~\ref{sec:examps}. Further, we will show that the ppE parameter for the total phase of the waveform is not necessarily that of the SPA phase given above, as is true for spin-aligned binaries.

The precession phase $\Phi_{K}^{\rm P}$ is generically given in Eq.~\eqref{eq:prec-phase}, and depends on the nutation phase $\psi$, Thomas phase $\epsilon$, and precession angle $\alpha$. The first of these is a purely secular function that evolves on the radiation-reaction timescale, while the former two are oscillatory on the precession timescale. The computation within each of these takes into account the near zone, far zone, and precession effects detailed above. As a result, one may specify
\begin{align}
    \label{eq:psi-nogr}
    \psi(u) &= \psi_{\rm GR}(u) + \mathscr{c} u^{c}
    \\
    \epsilon(u) &= \epsilon_{\rm GR}(u) + \mathscr{d} \delta\epsilon(u;\mathscr{c})
    \\
    \alpha(u) &= \alpha_{\rm GR}(u) + \mathscr{e} \delta\alpha(u;\mathscr{c})
\end{align}
where $(\mathscr{c},\mathscr{d},\mathscr{e})$ are parameters that capture non-GR effects, $c$ is an integer\footnote{Note that here, and for the rest of the paper, $c$ should not be confused with the speed of light in vacuum, which we have set to unity at the start of our analysis.}, and $(\delta\epsilon,\delta\alpha)$ are functions of $u$. The GR sector of each phase can be found in Eqs.~\eqref{eq:psi-dcs}, \eqref{eq:eps-dcs}, and \eqref{eq:alpha-dcs} for dCS gravity by taking the limit $\bar{\zeta}_{2} \rightarrow 0$, or in Eqs.~\eqref{eq:nut-quad}, \eqref{eq:eps-quad}, and \eqref{eq:alpha-quad} for nonaxisymmetric quadrupoles by taking the limit $(\cq_{1}, \cq_{2}) \rightarrow 0$, respectively.

Unlike the other phases, $(\epsilon,\alpha)$ cannot be written in the standard PN form, but instead in terms of a function of $u$ and $\psi$. Thus, we may be inclined to write
\begin{align}
    \delta \epsilon(u) &= \sum_{n} \delta\epsilon_{n}(u) e^{in\psi_{\rm GR}(u)} e^{in\mathscr{c} u^{c}}
    \\
    \delta \alpha(u) &= \sum_{n} \delta \alpha_{n}(u) e^{in\psi_{\rm GR}(u)} e^{in\mathscr{c} u^{c}}
\end{align}
In the case of dCS gravity, the oscillatory corrections appear by going to higher order in the MSA that was carried out in~\cite{Loutrel:2022tbk}, while for the nonaxisymmetric quadrupole scenario, the oscillatory effects can be solved for exactly. In either scenario, the oscillatory effects are actually PN suppressed, due to the fact that higher order terms in MSA are suppressed by the ratio of the precession timescale to the radiation-reaction timescale (in dCS gravity), and due to the fact that the secular behavior scales as $u^{-1}$ (in the case of deformed objects). Thus, we expect that the lowest order non-GR terms are those coming from the secular dynamics, and we write
\begin{align}
    \label{eq:eps-nogr}
    \epsilon(u) &= \epsilon_{\rm GR}(u) + \mathscr{d} u^{d}
    \\
    \label{eq:alpha-nogr}
    \alpha(u) &= \alpha_{\rm GR}(u) + \mathscr{e} \delta\alpha_{0}(u)
\end{align}
We are now left with determining the function $\delta \alpha_{0}(u)$.

In the non-axisymmetric quadrupole scenario, $\delta \alpha_{0}(u) \sim u^{-1}$ from Eq.~\eqref{eq:alpha-quad}. In dCS gravity, the correction is controlled by the functions $\varphi_{n}(u)$ given in Eq.~\eqref{eq:phin}, as opposed to a power of $u$. For example,
\begin{align}
    \varphi_{-3}(u) &= \frac{J_{0}(u)}{24 j_{0}^{2} u^{2}} \left[-8j_{0}^{2} + 3 j_{1}^{2} u^{2} - 2 j_{0} u \left(j_{1} + 4 j_{2} u\right)\right] 
    \nn \\
    &+ \frac{j_{1}}{8 j_{0}^{5/2}} \left(j_{1}^{2} - 4 j_{0} j_{2}\right) \tau_{1}
\end{align}
where the $j_{n}$ coefficients are given in Eq.~\eqref{eq:j-coeffs}, and 
\begin{equation}
    \tau_{1} = \tanh^{-1}\left\{\frac{u}{\sqrt{j_{0}}} \left[\sqrt{j_{2}} - J_{0}(u)\right]\right\}
\end{equation}
For other values of $n$, the expressions are given in Appendix D of~\cite{Loutrel:2022tbk}. Na\"ively, it does not appear that there is a generic function that one can write to eliminate the arbitrariness in $\delta \alpha_{0}(u)$. To work around this, we PN expand the functions $\varphi_{n}$ of dCS gravity and find
\begin{align}
    \varphi_{n}(u) &= \begin{cases}
        \frac{\sqrt{j_{0}}}{n} u^{n} & \text{if} \; n \neq 0\\
        \sqrt{j_{0}} \ln u & \text{if} \; n = 0
    \end{cases}
\end{align}
In the case $n=0$ above, $\varphi_{0} \sim {\cal{O}}(u^{0})$, since $\ln u \sim {\cal{O}}(1)$. The full functions $\varphi_{n}$ were originally defined in~\cite{Chatziioannou:2014coa}, where it was shown that PN expanding these functions resulted in a loss of accuracy in the GR precessing waveforms. However, here this function is multiplied by a non-GR parameter that is assumed to be small, and thus we are not concerned with the accuracy of these functions for astrophysical scenarios. By PN expanding these functions \textit{only} for the non-GR part of the waveform, we can write
\begin{equation}
    \delta \alpha_{0}(u) = u^{e}\,.
\end{equation}
which eliminates the arbitrariness in $\delta \alpha_{0}$.

Once one has the precession phases written in the form above, one has already considered all of the near zone, far zone, and precession effects that go into them. To get the Fourier-domain waveform, $u$ in these quantities must be promoted to the SUA variable $u_{k}$. Much like the SPA phase, we may deduce from the review in Sec.~\ref{sec:rev} that the SUA variable can be written as
\begin{align}
    \label{eq:sua-ppE}
    u_{k}(\tilde{u}) &= u_{k}^{\rm GR}(\tilde{u}) + k \mathscr{g} \tilde{u}^{g}\,,
\end{align}
where $u_{k}^{\rm GR}$ is given by Eq.~\eqref{eq:sua-eq} and $\mathscr{g}$ is a beyond-GR parameter. When $u\rightarrow u_{k}$ in the precession phases, the ``GR" part of each phase in terms of $u$ will become corrected by the ppE parameter $\mathscr{g}$. We must then deduce whether the SUA correction is lower or higher PN order than those in Eqs.~\eqref{eq:psi-nogr}  and ~\eqref{eq:eps-nogr}-\eqref{eq:alpha-nogr}. 

For concreteness, let us consider the nutation phase $\psi$ in Eq.~\eqref{eq:psi-nogr}, which when expanded becomes\footnote{From here, we explicitly show the $\times$ symbol whenever a constant symbols (such as $\mathscr{c}$) multiplies a function of $u_{k}(f)$, so as not to confuse them with functions of $u_{k}(f)$.}
\begin{align}
    \psi(f) &= \psi_{\rm GR}\left[u_{k}^{\rm GR}(f)\right] + \mathscr{c} \times \left[u_{k}^{\rm GR}(f)\right]^{c} 
    \nn \\
    &+ k \mathscr{g}_{\rm ppE} \times \left[u_{k}^{\rm GR}(f)\right]^{g_{\rm ppE}} \times \left[ \frac{\partial \psi_{\rm GR}}{\partial u}\right]_{u=u_{k}^{\rm GR}(f)}\,.
\end{align}
From the example cases in Sec.~\ref{sec:rev}, we know that $\psi$ is a purely secular function of $u$ and that it obeys a standard PN expansion. 
Thus, in analogy to Eqs.~\eqref{eq:psi-dcs} \& \eqref{eq:nut-quad}, we may take $\psi_{\rm GR} = \psi_{0}/u^{n_{\psi}} + {\cal{O}}(u^{-n_{\psi}+1})$, where $n_{\psi}$ controls the leading order of the PN expansion in GR, and $\psi_{0}$ is a known parameter. 
Thus, we have
\begin{align}
    \label{eq:psi-sua}
    \psi(f) &= \psi_{\rm GR}\left[u_{k}^{\rm GR}(f)\right] + \mathscr{c} \times \left[u_{k}^{\rm GR}(f)\right]^{c} 
    \nn \\
    &+ k \; \psi_{0} \; \mathscr{g} \times \left[u_{k}^{\rm GR}(f)\right]^{g-n_{\psi}-1}\,.
\end{align}
We now posit that $\psi$ can be written in the ppE-style form
\begin{align}
    \label{eq:psi-ppE}
    \psi_{\rm ppE}(f) &= \psi_{\rm GR}\left[u_{k}^{\rm GR}(f)\right] + \bar{\mathscr{c}} \times \left[u_{k}^{\rm GR}(f)\right]^{\bar{c}}\,,
\end{align}
and the problem reduces to determining the new parameters $(\bar{\mathscr{c}}, \bar{c})$ from Eq.~\eqref{eq:psi-sua}. In the spirit of the original ppE formalism, the new parameters are chosen from whichever of the latter two terms in Eq.~\eqref{eq:psi-sua} are lowest PN order. Thus, it follows that
\begin{align}
    \bar{c} &= \min\left(c, g - n_{\psi}-1\right)\,,
    \\
    \label{eq:c-bar}
    \bar{\mathscr{c}} &= \begin{cases}
        \mathscr{c} & \text{if} \; \bar{c} = c\\
        k \psi_{0} \mathscr{g} & \text{if} \; \bar{c} = g - n_{\psi}-1\\
        \mathscr{c} + k \psi_{0} \mathscr{g} & \text{if} \; \bar{c} = c = g - n_{\psi} -1
    \end{cases}
\end{align}
and we have fully specified the ppE parameters $(\bar{\mathscr{c}},\bar{c})$ for the nutation phase.

For the Thomas phase, the calculation is similar to what we presented above for $\psi$. Starting from Eq.~\eqref{eq:eps-nogr}, and promoting $u\rightarrow u_{k}$, we expand in small coupling to obtain
\begin{align}
    \epsilon(f) &= \epsilon_{\rm GR}\left[u_{k}^{\rm GR}(f)\right] + \mathscr{d} \times \left[u_{k}^{\rm GR}(f)\right]^{d} 
    \nn \\
    &+ k \mathscr{g} \times \left[u_{k}^{\rm GR}(f)\right]^{g} \times \left[ \frac{\partial \epsilon_{\rm GR}}{\partial u}\right]_{u=u_{k}^{\rm GR}(f)}\,.
\end{align}
Writing $\epsilon_{\rm GR} = \epsilon_{0}/u^{n_{\epsilon}} + {\cal{O}}(u^{-n_{\epsilon}+1})$, we can then posit the ppE formulation of $\epsilon$,
\begin{align}
    \label{eq:eps-ppE}
    \epsilon(f) &= \epsilon_{\rm GR}\left[\tilde{u}_{k}^{\rm GR}(f)\right] + \bar{\mathscr{d}} \times \left[u_{k}^{\rm GR}(f)\right]^{\bar{d}}
\end{align}
where
\begin{align}
    \bar{d} &= \min\left(d,g-n_{\epsilon}-1\right)\,,
    \\
    \label{eq:d-ppE}
    \bar{\mathscr{d}} &= \begin{cases}
        \mathscr{d} & \text{if} \; \bar{d} = d\\
        k \epsilon_{0} \mathscr{g} & \text{if} \; \bar{d} = g-n_{\epsilon}-1\\
        \mathscr{d} + k \epsilon_{0} \mathscr{g} & \text{if} \; \bar{d} = d = g - n_{\epsilon} - 1
    \end{cases}\,.
\end{align}
Likewise, for the precession angle we have
\begin{align}
    \alpha(f) &= \alpha_{\rm GR}\left[u_{k}^{\rm GR}(f)\right] + \mathscr{e} \times \left[u_{k}^{\rm GR}(f)\right]^{e} 
    \nn \\
    &+ k \mathscr{g} \times \left[u_{k}^{\rm GR}(f)\right]^{g} \times \left[ \frac{\partial \alpha_{\rm GR}}{\partial u}\right]_{u=u_{k}^{\rm GR}(f)}\,.
\end{align}
Writing $\alpha_{\rm GR} = \alpha_{0}/u^{n_{\alpha}} + {\cal{O}}(u^{-n_{\alpha}+1})$ for the derivative term above, we arrive at
\begin{align}
    \label{eq:alpha-ppE}
    \alpha(f) &= \alpha_{\rm GR}[u_{k}^{\rm GR}(f)] + \bar{\mathscr{e}} \times \left[u_{k}^{\rm GR}(f)\right]^{\bar{e}}
\end{align}
with
\allowdisplaybreaks[4]
\begin{align}
    \bar{e} &= \min\left(e,g - n_{\alpha} - 1\right)
    \\
    \label{eq:e-ppE}
    \bar{\mathscr{e}} &= \begin{cases}
        \mathscr{e} & \text{if} \; \bar{e} = e\\
        k\alpha_{0}\mathscr{g} & \text{if} \; \bar{e} = g - n_{\alpha} - 1\\
        \mathscr{e} + k \alpha_{0} \mathscr{g} & \text{if} \; \bar{e} = e = g - n_{\alpha} - 1
    \end{cases}\,.
\end{align}

Having now specified all of the precession angles in a ppE-style decomposition, we can write the ppE expansion of the precession phase $\delta \Phi^{\rm P}_{K}$. The full expression was given in complete generality in Eq.~\eqref{eq:Phi-decomp}, but from Eqs.~\eqref{eq:psi-ppE}, \eqref{eq:eps-ppE}, and~\eqref{eq:alpha-ppE}, we can write this as
\begin{align}
    \delta \Phi^{\rm P}_{K} &= n \bar{\mathscr{c}} \times \left[u_{k}^{\rm GR}(f)\right]^{\bar{c}} + m' \bar{\mathscr{e}} \times \left[u_{k}^{\rm GR}(f)\right]^{\bar{e}} 
    \nn \\
    &+ m \bar{\mathscr{d}} \times \left[u_{k}^{\rm GR}(f)\right]^{\bar{d}}\,.
\end{align}
Thus, we posit the ppE form,
\begin{equation}
    \label{eq:dPhiK}
    \delta \Phi_{K}^{\rm P} = \mathscr{c}^{\rm P}_{K} \times \left[u_{k}^{\rm GR}(f)\right]^{c^{\rm P}_{K}}
\end{equation}
where
\begin{align}
    c^{\rm P}_{K} &= \min\left(\bar{c},\bar{d},\bar{e}\right)\,,
    \\
    \label{eq:c-ppE}
    \mathscr{c}^{\rm P}_{K} &= \begin{cases}
        n \bar{\mathscr{c}} & \text{if} \; c_{K}^{\rm P} = \bar{c}\\
        m \bar{\mathscr{d}} & \text{if} \; c_{K}^{\rm P} = \bar{d}\\
        m' \bar{\mathscr{e}} & \text{if} \; c_{K}^{\rm P} = \bar{e}\\
        n \bar{\mathscr{c}} + m \bar{\mathscr{d}} & \text{if} \; c_{K}^{\rm P} = \bar{c} = \bar{d}\\
        n \bar{\mathscr{c}} + m' \bar{\mathscr{e}} & \text{if} \; c_{K}^{\rm P} = \bar{c} = \bar{e}\\
        m \bar{\mathscr{d}} + m' \bar{\mathscr{e}} & \text{if} \; c_{K}^{\rm P} = \bar{d} = \bar{e}\\
        n \bar{\mathscr{c}} + m \bar{\mathscr{d}} + m' \bar{\mathscr{e}} & \text{if} \; c_{K}^{\rm P} = \bar{c} = \bar{d} = \bar{e}
    \end{cases}\,.
\end{align}
The total phase of the waveform is the sum of the SPA phase and precession phase, and thus the correction is $\delta \tilde{\Psi}_{\rm tot} = \mathscr{b} \tilde{u}^{b} + \delta \Phi^{\rm P}_{K}$. Since the SUA function given in Eq.~\eqref{eq:sua-eq} is $u_{k}^{\rm GR} \sim \tilde{u} + {\cal{O}}(\tilde{u}^{7/2})$, we can employ an effective PN counting to determine the correction to the total phase of the waveform. We thus postulate
\begin{equation}
    \label{eq:ppE-phase}
    \delta \tilde{\Psi}_{\rm tot}^{\rm ppE} = \mathscr{b}^{\rm ppE}_{K} \times \left[U^{\rm ppE}_{K}(f)\right]^{b^{\rm ppE}_{K}}
\end{equation}
where
\begin{align}
    b_{K}^{\rm ppE} &= \min(b, c_{K}^{\rm P})\,,
    \\
    \mathscr{b}_{K}^{\rm ppE} &= \begin{cases}
        \mathscr{b} & \text{if} \; b_{K}^{\rm ppE} = b\\
        \mathscr{c}_{K}^{\rm P} & \text{if} \; b_{K}^{\rm ppE} = c_{K}^{\rm P}
    \end{cases}\,,
    \\
    U_{\rm ppE}(f) &= \begin{cases}
        \tilde{u} & \text{if} \; b_{K}^{\rm P} = b\\
        u_{k}^{\rm GR}(f) & \text{if} \; b_{K}^{\rm ppE} = c_{K}^{\rm P}
    \end{cases}\,.
\end{align}
Note that, in theory, one could also have a case where $b = c_{K}^{\rm P}$, in which case one would have to sum over both contributions. In such a case, the above mapping cannot be achieved, and one has instead
\begin{equation}
    \label{eq:ppE-phase-2}
    \delta \tilde{\Psi}_{\rm tot}^{\rm ppE} = \mathscr{b} \tilde{u}^{b} + \mathscr{c}_{K}^{\rm P} \times \left[u_{k}^{\rm GR}(f)\right]^{c_{K}^{\rm ppE}}
\end{equation}
In addition, the above ppE corrections are dependent on the harmonic numbers $(m,m',n)$, and thus, in principle, one could have corrections to multiple harmonics. We discuss how to choose which harmonics the ppE corrections should be added to in Sec.~\ref{sec:amp-ppE}.

\subsubsection{Amplitudes}
\label{sec:amp-ppE}

The amplitudes are given in terms of the Fourier decomposition of the SALPs by Eq.~\eqref{eq:amp-H}. When introducing beyond-GR effects, the Fourier coefficients $P_{K}(u)$ in the time-domain are shifted from their GR values due to the modification from the precession dynamics. These are generally the leading PN order corrections, since amplitude corrections from higher PN order radiative multipoles are PN suppressed. Is this the case for all beyond-GR scenarios? We postulate that this is indeed the case based on the following observation. In GR, there are only $l\ge2$ modes in the metric perturbation due to the fact that the $l=0$ and $l=1$ modes are forbidden by mass-energy and momentum conservation~\cite{Will:2014}. The same holds true in other modified theories of gravity, with one exception being those theories allowing for additional GW polarization states~\cite{Will:2014}. These additional states are not included in Eq.~\eqref{eq:h-def}, and must be handled separately. However, the procedure for doing so is simply the standard SUA procedure in GR. For the plus and cross polarizations, there are no $l<2$ modes and thus Eq.~\eqref{eq:h-def} holds.

To obtain the Fourier domain amplitudes for generic precessing waveforms, one has to take the time-domain $P_{K}(u)$ and promote $u \rightarrow u_{k}(f)$. For beyond-GR scenarios, this will introduce an additional deformation, besides the shifts
\begin{align}
    \label{eq:dPK-def}
    \delta P_{K}(u) &= \zeta^{A} \times \frac{\partial P_{K}(u;\zeta^{A})}{\partial \zeta^{A}}\,,
\end{align}
which are given in Eq.~\eqref{eq:dPK-dcs} for dCS gravity and Eq.~\eqref{eq:dPK-q} for non-axisymmetric quadrupoles. Using Eq.~\eqref{eq:sua-ppE} for the SUA function $u_{k}(f)$, we have the Fourier domain functions
\begin{align}
    \label{eq:PK-expand}
    P_{K}(f) &= P_{K}^{\rm GR}\left[u_{k}^{\rm GR}(f)\right]\left\{1 + \zeta^{A} \times \left[\frac{\partial \ln P_{K}}{\partial\zeta^{A}}\right]_{u=u_{k}^{\rm GR}}
    \right.
    \nn \\
    &\left.
    + k \mathscr{g} \times \left[u_{k}^{\rm GR}(f)\right]^{g} \times \left[ \frac{\partial \ln P_{K}^{\rm GR}}{\partial u}\right]_{u=u_{k}^{\rm GR}}\right\}
\end{align}
and we must determine whether the precession corrections coming from the second term above, or the SUA correction coming from the last term are the dominant PN contribution. Much like the case of the precession angle $\alpha$, the $P_{K}$ functions do not generally admit a typical PN expansion, but are instead complicated functions of $u$. We again employ a PN expansion of only the non-GR part of the above expression, and take $(n_{1},n_{2})$ to be the effective PN orders of the latter two terms in Eq.~\eqref{eq:PK-expand}. Then, we can write
\begin{equation}
    P_{K}^{\rm ppE}(f) = P_{K}^{\rm GR}\left[u_{k}^{\rm GR}(f)\right] \times \left\{1 + \mathscr{a}_{K}^{\rm ppE} \times \left[u_{k}^{\rm GR}(f)\right]^{a_{K}^{\rm ppE}}\right\}
\end{equation}
with
\allowdisplaybreaks[4]
\begin{align}
    a_{K}^{\rm ppE} &= \min(n_{1}, g + n_{2})
    \\
    \label{eq:amp-ppE}
    \mathscr{a}_{K}^{\rm ppE} &= \begin{cases}
        \zeta^{A} \mathscr{p}_{1} & \text{if} \; p=n_{1}\\
        k \mathscr{g} \mathscr{p}_{2} & \text{if} \; p=g + n_{2}\\
        \zeta^{A} \mathscr{p}_{1} + k \mathscr{g} \mathscr{p}_{2} & \text{if} \; p = n_{1} = g + n_{2}
    \end{cases}\,,
\end{align}
and where $[\mathscr{p}_{1,2},n_{1,2}]$ are defined such that
\begin{align}
    \frac{\partial \ln P_{K}}{\partial \zeta^{A}} &= \mathscr{p}_{1} u^{n_{1}} + {\cal{O}}(u^{n_{1}+1})\,,
    \\
    \frac{\partial \ln P_{K}^{\rm GR}}{\partial u} &= \mathscr{p}_{2} u^{n_{2}} + {\cal{O}}(u^{n_{2}+1})\,.
\end{align}

At this stage, ppE deformations will exist in every harmonic of the waveform, which is in contrast to the original ppE framework on non-precessing binaries where the deformations only appear in the dominant harmonic. This creates a problem if one wants to perform theory agnostic tests of GR using this new framework, since having deformations in multiple harmonics can worsen constraints on any one deformation. To address this, we will make use of some data analysis considerations in Sec.~\ref{sec:da}, which will allow us to limit the new ppE framework to only include deformation in a subset of harmonics, labeled $\slashed{K}$. For instructive purposes, we provide the final mappings for the example scenarios first in Sec.~\ref{sec:examps} and leave the technical details to Sec.~\ref{sec:da}.

\subsubsection{The Precessing ppE Waveform}
\label{sec:ppE-wav}

Combining all of the above considerations together, we postulate the following precessing ppE waveform,
\begin{align}
    \label{eq:prec-ppE}
    \tilde{h}(f) &= \sum_{\tilde{K}} \tilde{h}^{\rm GR}_{\tilde{K}}(f) e^{i \mathscr{b}^{\rm ppE}_{\slashed{K}} \times \left[U^{\rm ppE}_{\slashed{K}}(f) \right]^{b^{\rm ppE}_{\slashed{K}}}}
    \nn \\
    &\times \left\{1 + \mathscr{a}^{\rm ppE}_{\slashed{K}} \times \left[\tilde{u}_{k}^{\rm GR}(f)\right]^{a^{\rm ppE}_{\slashed{K}}} \right\}
\end{align}
where $\tilde{u} = (2\pi M f/|m|)^{1/3}$,
\begin{align}
    \tilde{h}_{\rm GR} &= {\cal{A}}_{\tilde{K}}^{\rm GR}(f) e^{i[\tilde{\Psi}_{m}^{\rm GR}(f) + \Phi^{\rm P,GR}_{mm'n}(f)]}\,,
\end{align}
with $\tilde{\Psi}_{m}(f)$ given by Eq.~\eqref{eq:spa-def}, $\Phi^{\rm P,GR}_{mm'n}$ given by Eq.~\eqref{eq:prec-phase},
\begin{align}
    {\cal{A}}_{\tilde{K}}^{\rm GR}(f) &= -\sqrt{\frac{2}{3}} \frac{{\cal{M}}^{5/6}N_{lm}}{D_{L} \pi^{1/6}f^{7/6}}  \left(\frac{a_{k, k_{\rm max}}}{2 - \delta_{k,0}}\right)  P_{\tilde{K}}^{\rm GR}\left[\tilde{u}_{k}^{\rm GR}(f)\right]\,,
\end{align}
$\tilde{K} = K \cup k = lmm'nk$ is the hyper-index such that
\begin{align}
    \sum_{\tilde{K}} &\rightarrow \sum_{l\ge2}\sum_{m<0}\sum_{m'=-l}^{l}\sum_{n=-\infty}^{\infty}\sum_{k=-k_{\rm max}}^{k_{\rm max}}\,,
\end{align}
and $\slashed{K}$ denotes the subset of harmonics $K$ which possess ppE deformations. For spin-aligned, quasi-circular binaries, the ppE waveform is given in Eq.~\eqref{eq:ppE}, with four total parameters $(\alpha_{\rm ppE}, \beta_{\rm ppE}, a_{\rm ppE}, b_{\rm ppE})$. 
In contrast, the precessing ppE waveform is parameterized by two sets of parameters $[\mathscr{b}_{\slashed{K}}^{\rm ppE}, b_{\slashed{K}}^{\rm ppE}]$ for the total phase, and two sets of parameters $[\mathscr{a}_{\slashed{K}}^{\rm ppE}, a_{\slashed{K}}^{\rm ppE}]$ for the waveform amplitudes.

\subsection{Example Mappings}
\label{sec:examps}

We now provide the mappings between the precessing ppE parameters and beyond-GR parameters of the scenarios considered in Sec.~\ref{sec:rev}. The full technical details of why each harmonics is chosen to posses ppE deformations are provided in Sec.~\ref{sec:da}. We provide the example mappings here for instructive purposes, since it is useful to know the final answer before getting into the full details of the analysis. For both of the example cases $l=2$ and $m=-2$, so we only need to consider variations over the harmonics numbers $(m',n)$.

\subsubsection{dCS gravity}
\label{sec:ppE-dcs}

In dCS gravity, the corrections to the SPA phase are given by Eqs.~\eqref{eq:spa-dcs-1}-\eqref{eq:spa-dcs-2}, while the SPA phase in the ppE formalism is given by Eq.~\eqref{eq:spa-ppE}. The map between these are
\begin{align}
    \mathscr{b} &= 4 \bar{\zeta}_{2} \langle \delta \beta_{3}^{\rm dCS}\rangle_{\psi}\,, \qquad b = -2\,.
\end{align}
For the precession phase, the SUA correction enters at 3.25PN order in Eq.~\eqref{eq:sua-eq} such that $g=13/2$ in Eq.~\eqref{eq:sua-ppE}. Meanwhile, the correction to the nutation phase in Eq.~\eqref{eq:psi-dcs} enters at -0.5PN order such that $c=-1$ in Eq.~\eqref{eq:psi-nogr}. For the spin precession problem in GR, $n_{\psi} = 4$. Since $g_{\rm ppE}-n_{\psi}=5/2$ which is greater than $c$, we have
\begin{align}
    \bar{\mathscr{c}} &= -\frac{5}{128} \frac{(1-q^{2})}{q} \delta \psi_{2} \bar{\zeta}_{2}\,,
    \qquad
    \bar{c} = -1\,.
\end{align}
Following this same analysis for $\epsilon$ and $\alpha$, we have
\begin{align}
    \label{eq:dcs-d}
    \bar{\mathscr{d}} &= \bar{\zeta}_{2} \delta \epsilon_{-3}^{(-1)}\,,
    \qquad 
    \bar{d} = -3\,,
    \\
    \label{eq:dcs-e}
    \bar{\mathscr{e}} &= -\frac{\bar{\zeta}_{2}}{3} \sqrt{j_{0}} \delta \alpha_{-3}^{(-1)}\,,
    \qquad
    \bar{e} = -3\,.
\end{align}
Note that the ppE correction for $\alpha$ here only contains the first term in the brackets in Eq.~\eqref{eq:alpha-dcs}. The reason for this is that, due to the dependence on $J$, $\delta \varphi_{n}(u)$ are higher effective PN order than $\varphi_{n}(u)$ (see Eq.~\eqref{eq:dphi-3}). From these considerations, it is clear that the corrections to the Thomas phase and precession angle are more important than the corrections to the nutation phase, since they are lower PN order.

Based on the analysis in Sec.~\ref{sec:da}, the most important harmonic for dCS gravity are the $(m',n) = (+1,0)$, $(+1,+2)$, and $(+2,0)$ harmonics. For the waveform phase, the ppE parameters are
\begin{align}
    \label{eq:b-dcs}
    \mathscr{b}_{(+1,0)}^{\rm ppE} &= \mathscr{b}_{(+1,+2)}^{\rm ppE} = \bar{\zeta}_{2} \left[-2 \delta \epsilon_{-3}^{(-1)} - \frac{1}{3} \sqrt{j_{0}} \delta \alpha_{-3}^{(-1)}\right]\,,
    \nn \\
    b_{(+1,0)}^{\rm ppE} &=  b_{(+1,+2)}^{\rm ppE} = -3\,,
    \\
    \mathscr{b}_{(+2,0)}^{\rm ppE} &= 4 \bar{\zeta}_{2} \langle \delta \beta_{3} \rangle_{\psi}\,, \qquad b_{(+2,0)}^{\rm ppE} = -2\,,
\end{align}
and the functions $U^{\rm ppE}_{\slashed{K}}(f)$ are
\begin{align}
    U^{\rm ppE}_{(+1,0)} &= U^{\rm ppE}_{(+1,+2)} = u_{k}^{\rm GR}(f)\,, \qquad U^{\rm ppE}_{(+2,0)} &= \tilde{u}\,.
\end{align}
For the $(+2,0)$ harmonic, the contribution from $\bar{d}$ cancels the contribution from $\bar{e}$ in Eqs.~\eqref{eq:dcs-d} \&~\eqref{eq:dcs-e}, respectively. This cancellation also happens at next PN order, and thus the correction enters at $-1.5$ PN order, specifically arising from the SPA phase. The contribution from the nutation phase is PN suppressed.

Finally, the amplitude ppE parameters are
\begin{widetext}
\begin{align}
    \mathscr{a}_{(+1,0)}^{\rm ppE} &= \frac{s_{+}^{(0)} \left(2 \delta s_{-}^{(0)} + 3 \delta s_{+}^{(0)}\right) - 2 c_{1}^{2} \left(\delta s_{-}^{(0)} + \delta s_{+}^{(0)}\right) - s_{-}^{(0)} \delta s_{+}^{(0)}}{2 \left(4 c_{1}^{2} - s_{-}^{(0)} - 3 s_{+}^{(0)}\right) \left(c_{1}^{2} - s_{+}^{(0)}\right)}\,, \qquad a_{(+1,0)}^{\rm ppE} = 0\,,
    \\
    \mathscr{a}_{(+1,+2)}^{\rm ppE} &= \frac{s_{+}^{(0)} \left(2 \delta s_{-}^{(0)} - \delta s_{+}^{(0)}\right) - 2 c_{1}^{2} \left(\delta s_{-}^{(0)} - \delta s_{+}^{(0)}\right) - s_{-}^{(0)} \delta s_{+}^{(0)}}{2 \left(s_{+}^{(0)} - s_{-}^{(0)}\right) \left(c_{1}^{2} - s_{+}^{(0)}\right)}\,, \qquad a_{(+1,+2)}^{\rm ppE} = 0\,,
    \\
    \label{eq:a-dcs}
    \mathscr{a}_{(+2,0)}^{\rm ppE} &= -\frac{\left(\delta s_{+}^{(0)} + \delta s_{-}^{(0)}\right)}{4 M^{4} \eta^{2}}\,, \qquad a_{(+2,0)}^{\rm ppE} = 2\,.
\end{align}
\end{widetext}
Note that while $a_{(+1,0)}^{\rm ppE} = 0 = a_{(+1,+2)}^{\rm ppE}$, this does not mean that these harmonics do not evolve in time (frequency). This simply means that these corrections appear at the same PN order as the GR sector of these harmonics, i.e. they are relative Newtonian order corrections. For convenience, we point out that in this case the SALPs ${_{m'}}P_{2,-2}$ in the the GR sector scale as $u^{|m'-2|}$. This completes the mapping of the precessing ppE waveform to dCS gravity.

\subsubsection{Non-axisymmetric quadrupoles}
\label{sec:ppE-quad}

The nonaxisymmetric quadrupole scenario differs significantly from the dCS scenario. The latter is a perturbation of the spin-precession problem in GR, while the former are the perturbations to the precession dynamics for nonspinning bodies with spheroidal quadrupole moments. The first main difference is that the quadrupole case is described by two beyond-GR parameters that are considered to be small, specifically $\cq_{1}$ and $\cq_{2}$. Thus, the ppE parameters for the SPA phase are
\begin{align}
    \label{eq:spa-quad}
    \mathscr{b} &= \frac{5}{4}\sqrt{\frac{\pi}{5}} \chi_{Q} \left[\cq_{1} {\cal{U}}_{10} + \cq_{2} {\cal{U}}_{01}\right]\,, \qquad b = -1\,.
\end{align}
The second main difference from the dCS case is that the nutation phase is different. For this scenario, the nutation phase is $\tilde{\psi}$ given in Eq.~\eqref{eq:nut-quad}, with the leading-order behavior for both GR and non-GR scales as $u^{-1}$, and thus $n_{\psi} = c = -1$. The SUA correction is given in Eq.~\eqref{eq:sua-q} and enters at 3.75PN order, i.e. $g_{\rm ppE} = 15/2$. Once again, $g_{\rm ppE} - n_{\psi} = 13/2$ which is greater than $c$. Thus, the deformation parameters for the nutation phase is
\begin{align}
    \bar{\mathscr{c}} &= \frac{3\sqrt{5\pi}}{32} \chi_{Q} \Gamma(\cq_{1},\cq_{2})\,,\qquad \bar{c} = -1\,,
\end{align}
where
\begin{align}
    \Gamma(\cq_{1},\cq_{2}) &= \cq_{1} \tan\beta_{0} \sin\Delta + \frac{1}{4} \cq_{2} \tan^{2}\beta_{0}\,.
\end{align}
Following the same arguments, the corrections for the Thomas phase and precession angle are
\begin{align}
    \bar{\mathscr{d}} &= \frac{3\sqrt{5\pi}}{32} \chi_{Q} \Omega_{\rm \epsilon} \Gamma(\cq_{1},\cq_{2})\,, \qquad \bar{d} = -1\,,
    \\
    \bar{\mathscr{e}} &= - \frac{3\sqrt{5\pi}}{32} \chi_{Q} \Gamma(\cq_{1},\cq_{2})\,, \qquad \bar{e} = -1\,.
\end{align}
Note that in this case, all of the phase corrections are degenerate, i.e. they all enter at the same PN order.

From the analysis in Sec.~\ref{sec:da}, the most important harmonics for the non-axisymmetric quadrupole case are $(m',n) = (0,0)$, $(+1,0)$, $(+2,0)$, and $(-2,0)$. We thus have the following ppE parameters for the non-axisymmetric quadrupole scenario,
\begin{widetext}
\begin{align}
    \label{eq:phase-quad-ppE}
    \delta \tilde{\Psi}_{\rm tot}^{\rm ppE} &= \mathscr{b} \tilde{u}^{b} + \mathscr{c}_{\slashed{K}}^{\rm ppE} \times \left[u_{k}^{\rm GR}(f)\right]^{c_{\slashed{K}}^{\rm P}}\,,
    \\
    \mathscr{c}_{(m',n)}^{\rm P} &= \frac{3\sqrt{5\pi}}{32} \left(n - m' - 2\right) \chi_{Q} \Gamma(\mathscr{q}_{1}, \mathscr{q}_{2})\,, \qquad c_{(m',n)}^{\rm P} = -1\,,
    \\
    \mathscr{a}_{(0,0)}^{\rm ppE} &= - \mathscr{q}_{2} + 2 \mathscr{q_{1}} \cot\beta_{0} \sin \Delta\,, \qquad a_{(0,0)}^{\rm ppE} = 0\,,
    \\
    \mathscr{a}_{(+1,0)}^{\rm ppE} &= 2 \mathscr{q}_{1} \left(2 \cos\beta_{0} - 1\right) \csc\beta_{0} \sin\Delta - \mathscr{q}_{2} \tan\beta_{0}\,, \qquad a_{(+1,0)}^{\rm ppE} = 0\,,
    \\
    \mathscr{a}_{(+2,0)}^{\rm ppE} &= \tan\left(\frac{\beta_{0}}{2}\right) \left(\mathscr{q}_{2} \tan\beta_{0} - 2 \mathscr{q}_{1} \sin\Delta\right)\,, \qquad a_{(+2,0)}^{\rm ppE} = 0\,,
    \\
    \label{eq:amp20-quad-ppE}
    \mathscr{a}_{(-2,0)}^{\rm ppE} &= \cot\left(\frac{\beta_{0}}{2}\right) \left(2 \mathscr{q}_{1} \sin\Delta - \mathscr{q}_{2} \tan\beta_{0}\right)\,, \qquad a_{(-2,0)}^{\rm ppE} = 0\,,
\end{align}
\end{widetext}
where $[\mathscr{b}, b]$ are given in Eq.~\eqref{eq:spa-quad}. Since the corrections to the precession phase are all degenerate, the ppE deformation is characterized by a superposition of the corrections to the nutation phase, Thomas phase, precession angle, and SPA phase. The amplitude corrections all enter at relative Newtonian order, and thus $a_{(m',n)}^{\rm ppE} = 0$. However, unlike the case of dCS gravity, the GR sector of the SALPs is not time (frequency) dependent. This is due to the fact that the non-axisymmetric deviations are considered perturbations of a spheroidal configuration, for which the inclination angle is a constant. 

\subsection{Toward a Minimal Construction Through Data Analysis Considerations}
\label{sec:da}

Having provided the final mapping of the ppE parameters to the example cases, we now provide the details of how we arrived at these mappings. The analysis follows basic data analysis considerations, particularly, a calculation of the likelihood between GR and non-GR waveforms to determine the importance of each harmonic, and the overlap as a test of whether the mapping in the previous section are faithful.

\subsubsection{Likelihood Considerations}
\label{sec:like}

How does one determine what is the most important harmonic to attach the ppE corrections to? When only one harmonic is present, one should simply choose the lowest PN order correction to the amplitude. However, if there are multiple harmonics, it is possible that corrections to two (or more) harmonics appear at the same PN order. To determine which harmonic is most important, we investigate the likelihood function, which defines the noise model. For stationary and Gaussian noise, the likelihood function will therefore tell us how ``close'' a signal is to a model for a given set of parameters. If the signal is the full GR model with many harmonics and the model is a  modified gravity model, then minimizing the log likelihood with respect to all parameters will tell us how close to GR the modified gravity model is allowed to be given statistical uncertainties. If, on the other hand, the model is a deformation from GR, we can investigate the log-likelihood to determine which harmonic in the GR deformation contributes the most. 

Before carrying out such a study, however, let us establish some notation. Given two time-domain models $A(t)$ and $B(t)$, the noise-weighted inner product is defined as
\begin{align}
    (A | B) = 4 \Re \int_{f_{\rm low}}^{f_{\rm high}} \frac{df}{S_{n}(f)} \tilde{A}(f) \tilde{B}^{\dagger}(f)
\end{align}
where $\Re$ is the real part operator, the overhead tilde stands for the Fourier transform, the $\dagger$ is the complex conjugate, and $S_{n}(f)$ is the power spectral density of the detector. For simplicity and also to allow our analysis to be detector-agnostic, we set $S_{n}(f) = 1$. 

The waveforms for precessing binaries depend on the system orientation angles $(\theta_{N}, \phi_{N})$ in the following manner
\begin{align}
    \label{eq:sph-harm-decomp}
    \tilde{h}(f) = \sum_{lm'} \tilde{h}_{lm'}(f) {_{-2}}Y_{lm'}(\theta_{N}, \phi_{N})
\end{align}
Rather than consider systems across multiple different orientations, we define the averaged inner product as follows. Allow two waveforms $[\tilde{A}(f), \tilde{B}(f)]$ to be decomposed as in Eq.~\eqref{eq:sph-harm-decomp}. Then, the averaged inner product is
\begin{align}
    \langle A | B \rangle &= 4 \Re \int_{f_{\rm low}}^{f_{\rm high}} df \int d\Omega_{N} \tilde{A}(f,\theta_{N}, \phi_{N}) \tilde{B}^{\dagger}(f,\theta_{N}, \phi_{N})
    \nn \\
    &= 4 \Re \sum_{lm'} \int_{f_{\rm low}}^{f_{\rm high}} \tilde{A}_{lm'}(f) \tilde{B}^{\dagger}_{lm'}(f)
    \nn \\
    &= \sum_{lm'} (A_{lm'} | B_{lm'})
\end{align}
where $d\Omega_{N} = \sin\theta_{N} d\theta_{N} d\phi_{N}$, and we have made use of the orthogonality between spin-weighted spherical harmonics, i.e.
\begin{equation}
    \int d\Omega \; {_{s}}Y_{lm}(\theta, \phi) {_{s}}Y^{\dagger}_{l'm'}(\theta, \phi) = \delta_{ll'} \delta_{mm'}\,.
\end{equation}
We then define the averaged log-likelihood between two waveforms as
\begin{equation}
    \label{eq:like-def}
    \langle \ln {\cal{L}} \rangle = -\frac{1}{2} \langle A-B | A-B \rangle\,.
\end{equation}

As explained above, we here wish to use the averaged log-likelihood to determine which harmonic matters the most in the GR deformation model when carrying out tests of GR. Therefore, we choose the two waveforms to be as follows: the signal will be a precessing waveform in GR with many harmonics; the recovery model will be a GR deformation of the precessing GR waveform also with many harmonics. Let us then write the precessing GR waveform as
\begin{equation}
    \tilde{h}_{\rm GR}(f) = h_{0} f^{-7/6} \sum_{K} {\cal{A}}_{K}(f) e^{i\Psi_{K}} {_{-2}}Y_{lm'}(\theta_{N}, \phi_{N})\,.
\end{equation}
where $K = lmm'n$ is the multi-spectral index, and $h_{0}$ is an overall pre-factor that is independent of frequency. Here, $\Psi_{K}$ includes both the SPA phase and the precession phase, i.e. $\Psi_{K}(f) = \tilde{\Psi}_{m}(f) + \Phi_{lmm'n}^{\rm P}(f)$. Now, let us write the beyond-GR waveform as a deformation of the above waveform via 
\begin{align}
    \tilde{h}(f) &= h_{0} f^{-7/6} \sum_{K} \left[{\cal{A}}_{K}(f) + \zeta^{A} \delta {\cal{A}}_{K}\right] 
    \nn \\
    &\times e^{i\left[\Psi_{K}(f) + \zeta^{A} \delta \Psi_{K}(f)\right]} {_{-2}}Y_{lm'}(\theta_{N}, \phi_{N})
\end{align}
where $\zeta^{A}$ is a small deformation parameter. The log-likelihood calculation requires we compute the inner product of the difference between these two waveforms. Doing so and linearizing in $\zeta^{A}$ gives
\begin{align}
    \tilde{h}(f) - \tilde{h}_{\rm GR}(f) &= h_{0}f^{-7/6} \zeta^{A} \sum_{K} \left[\delta {\cal{A}}_{K}(f) + i {\cal{A}}_{K} \delta \Psi_{K}(f)\right]
    \nn \\
    &\times e^{i\Psi_{K}(f)} {_{-2}}Y_{lm'}(\theta_{N}, \phi_{N})
\end{align}
Thus, the averaged log-likelihood of Eq.~\eqref{eq:like-def} becomes
\begin{align}
    \label{eq:lnL-tot}
    \langle \ln {\cal{L}} \rangle &= 4 h_{0}^{2} (\zeta^{A})^{2} \sum_{L\cup L'} \langle \delta \ln  {\cal{L}} \rangle_{L \cup L'}
\end{align}
where we have defined
\begin{widetext}
\begin{align}
    \label{eq:lnL-all}
    \langle \delta \ln {\cal{L}} \rangle_{L\cup L'} &= -\frac{1}{2} \Re \int_{f_{\rm low}}^{f_{\rm high}} \frac{df}{f^{7/3}} \Bigg\{\delta {\cal{A}}_{L}(f) \delta {\cal{A}}_{L'}(f) + {\cal{A}}_{L}(f) {\cal{A}}_{L'}(f) \delta \Psi_{L}(f) \delta\Psi_{L'}(f)
    \nn \\
    &+ i \left[\delta A_{L'}(f) {\cal{A}}_{L}(f) \delta \Psi_{L}(f) - \delta {\cal{A}}_{L}(f) {\cal{A}}_{L'}(f) \delta\Psi_{L'}(f)  \right] \Bigg\} e^{i \Psi_{L}(f) - i \Psi_{L'}(f)}\,,
\end{align}
\end{widetext}
with $L = lm'm_{1}n_{1}$, $L'=lm'm_{2}n_{2}$, and $L\cup L' = lm'm_{1}m_{2}n_{1}n_{2}$ are multi-indices. Note that, due to the orientation averaging, $L\cup L'$ only contains single values of $(l,m')$. 

We have then arrived at a exression that gives us the averaged log-likelihood [Eq.~\eqref{eq:lnL-tot}] as a sum over harmonics of the averaged deformed log-likelihood. Each term in this sum can now be evaluated (term by term) to determine which harmonic contributes the most to the log-likelihood, and thus, which harmonics produce the largest deviations from GR. In general, one could compute the total log-likelihood in Eq.~\eqref{eq:lnL-tot}, but since we are comparing harmonic to harmonic, it is simpler to compute the harmonic likelihood in Eq.~\eqref{eq:lnL-all}, and take $L = L'$. In fact, if one considers the beyond-GR waveform to be a ppE-style waveform that only has one amplitude and one phase correction, then the sums above collapse and ensure that $L=L'$. In this scenario, Eq.~\eqref{eq:lnL-all} reduces to
\begin{align}
    \langle \delta \ln {\cal{L}} \rangle_{L} &= \lim_{L'\rightarrow L} \langle \delta \ln {\cal{L}} \rangle_{L\cup L'} 
    \nonumber \\
    &= -\frac{1}{2} \int_{f_{\rm low}}^{f_{\rm high}} \frac{df}{f^{7/3}} \Bigg\{\left[\delta {\cal{A}}_{L}(f)\right]^{2} 
    \nn \\
    &+ \left[{\cal{A}}_{L}(f)\right]^{2} \left[\delta \Psi_{L}(f)\right]^{2} \Bigg\}\,,
    \label{eq:lnL-harm}
\end{align}
where we have dropped the $\Re$ operator because the integral is now manifestly real valued, and $L$ is now $L=lm'mn$. For the two beyond-GR scenarios we consider here, $l=2$ and $m=-2$, so it suffices to only consider the $(m',n)$ harmonics. 

The integral in Eq.~\eqref{eq:lnL-harm} can be simplified by PN expanding the quantities $[\delta {\cal{A}}_{L}, {\cal{A}}_{L}, \delta \Psi_{L}]$, such that all quantities scale as $u^{p}$, where $p$ varies depending on the quantity and harmonic numbers considered. In general, one would have to promote $u$ to the SUA function $u_{k}^{\rm GR}$, but in practice, we find that this does not change the relative qualitative behavior of $\langle \delta \ln {\cal{L}}\rangle_{L}$ for different $L$, only its overall value. Further, the corrections in the SUA function $u_{k}^{\rm GR}$ in Eq.~\eqref{eq:sua-eq} scale as $\tilde{u}^{7/2}$, and are thus PN suppressed. Thus, in practice, we do not promote $\tilde{u}\rightarrow \tilde{u}_{k}^{\rm GR}$ when evaluating Eq.~\eqref{eq:lnL-harm}, and we simply use the mapping $\tilde{u} = (2\pi M f/|m|)^{1/3}$. Lastly, with the above assumptions, the amplitude functions $[{\cal{A}}_{L}, \delta {\cal{A}}_{L}]$ reduce to the Fourier coefficients of the SALPs, specifically $[P_{K}(u), \delta P_{K}(u)]$. 

The purpose of computing $\langle \delta \ln {\cal{L}}\rangle_{L}$ is to answer two questions, specifically:
\begin{itemize}
    \item Which $(m',n)$ harmonic produces the largest value of $\langle \delta \ln {\cal{L}}\rangle_{L}$? We refer to this harmonic as the \textit{maximum likelihood harmonic}~(MLH).
    \item When computing the total log-likelihood, how many harmonics are needed to obtain a desired value of the total, assuming the sum is performed over descending order in the value of $\langle \delta \ln{\cal{L}}\rangle_{L}$? We refer to this subset of harmonics as the \textit{requisite subset of harmonics}~(RSH). 
\end{itemize}
The answer to the first question tells us which is the most important harmonic, and na\"ively, one might expect to only include amplitude deviations to that one harmonic. However, in practice, we find that only including this one harmonic does not account for a sufficiently large amount of the total likelihood, and one has to sum over multiple harmonics to achieve a desired percentage of the total. For all of the cases studied here, we choose this percentage to be 90\%, but whether this cutoff is high enough needs to be determined with a careful Bayesian parameter estimation study, which we leave to future work. 
As a last point, when computing the sum in Eq.~\eqref{eq:lnL-tot}, we only perform the sum over $L=L'$. The terms in the sum with $L\neq L'$ are determined by Eq.~\eqref{eq:lnL-all}, which is a highly oscillatory integral, and are typically greatly suppressed.

\vspace{0.2cm}
\begin{center}
\noindent{\emph{a. Likelihood analysis of dCS gravity}}
\end{center}
\vspace{0.2cm}

Let us first consider the case of dCS gravity. We calculate Eq.~\eqref{eq:lnL-harm} across all possible combinations of harmonic numbers $(m',n)$. Note that due to the nutation dynamics, there are only even values for $n$. Further, we restrict our attention to only the $n=0$ and $n=\pm 2$ harmonics, since higher $n$ harmonics are suppressed by higher powers of the spins of the BHs. We perform the analysis for $10^{4}$ BBH systems, where the parameters are chosen randomly (with a flat prior). The ranges for the total mass and mass ratio are $M \in [10,40] M_{\odot}$ and $q\in[1/10,9/10]$. We do not consider the equal mass case due to subtleties in taking this limit in dCS gravity (see~\cite{Loutrel:2022tbk}). The dimensionless spins of the BHs are allowed to take all possible values for Kerr BHs, i.e. $\chi_{1,2} \in [0,1]$. The initial values of the total angular momentum $J$ and total spin angular momentum $S$ are chosen at random (and with a flat prior) from
\begin{align}
    J \in \left[J_{\rm min}, J_{\rm max}\right]\,, \qquad S \in \left[S_{\rm min}, S_{\rm max}\right]\,,
\end{align}
where
\begin{align}
    J_{\rm min} &= L - S_{1} - S_{2}\,, 
    \\
    J_{\rm max} &= L + S_{1} + S_{2}\,,
    \\
    S_{\rm min} &= \max\left(J-L, S_{1}-S_{2}\right)\,,
    \\
    S_{\rm max} &= \min\left(J+L, S_{1} + S_{2} \right)\,.
\end{align}
The value of the mass-weighted effective spin $\chi_{\rm eff}$ can be randomly chosen by using Eq.~(28) in~\cite{Loutrel:2022tbk} and considering that $\phi'\in[0,2\pi]$. The quantity $c_{1}$ can then be solved for by applying Eq.~\eqref{eq:J-dcs}.

\begin{figure*}[hbt]
    \centering
    \includegraphics[width=\textwidth, trim={2cm 0cm 4cm 0cm}, clip]{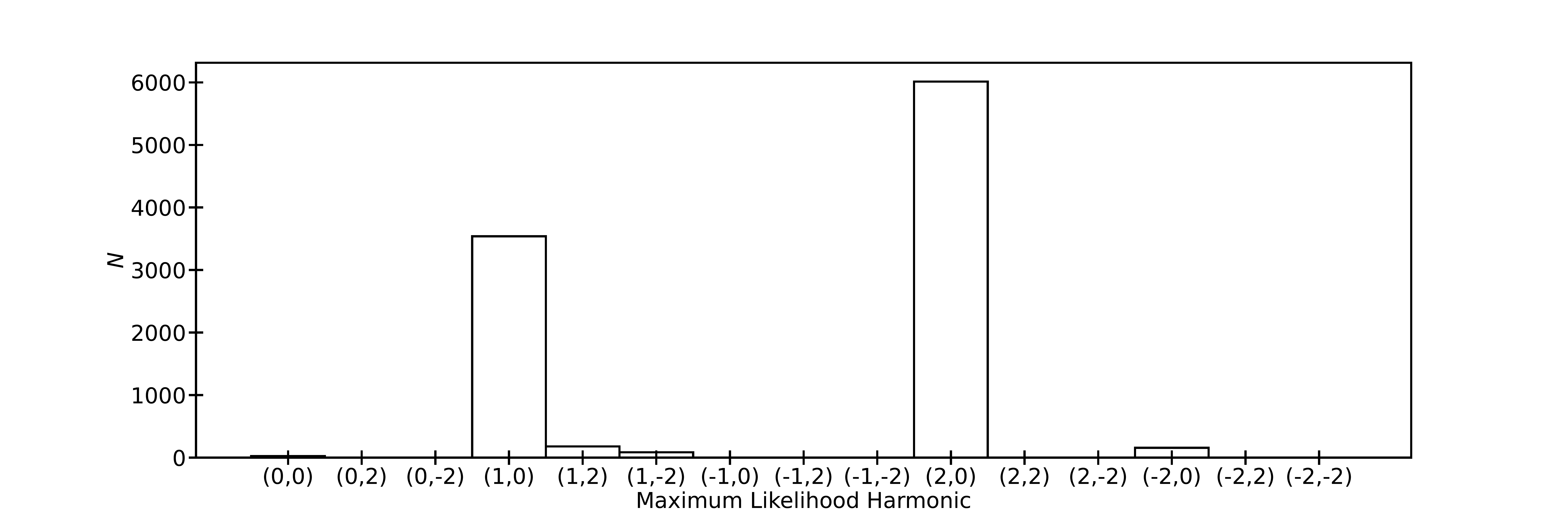}
    \includegraphics[width=\textwidth, trim={2cm 0cm 4cm 0cm}, clip]{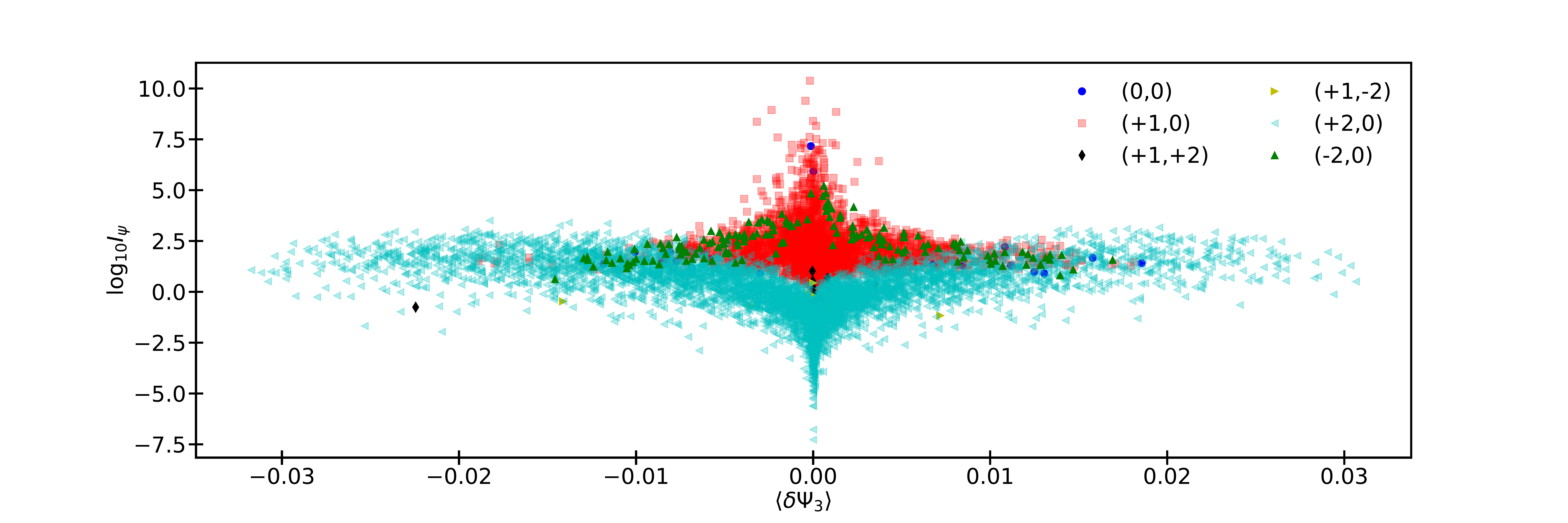}
    \caption{\label{fig:dcs} Top: Histogram of the MLH for each BBH system considered in dCS gravity. Observe that most systems have either the $(+1,0)$ or $(+2,0)$ harmonics as the MLH, with the other harmonics being outliers. Bottom: MLH as a function of the SPA spin-orbit correction $\langle \delta \Psi_{3}\rangle$ and the amplitude weighted phase difference $I_{\psi}$ defined in Eq.~\eqref{eq:Ipsi}.}
\end{figure*}

Let us now study which harmonic produces the largest $\langle \ln{\cal{L}} \rangle_{L}$, where here $L=(m',n)$. Figure~\ref{fig:dcs} shows the results of the MLH computation. For all of the $10^{4}$ BBH systems, the $(m',n)$ harmonics that contribute the most to the averaged deformed log-likelihood are $(0,0)$, $(+1,0)$, $(+1,+2)$, $(+1,-2)$, $(+2,0)$, or $(-2,0)$. The top panel of Fig.~\ref{fig:dcs} shows a histogram of the MLH. The overwhelming majority of systems have either the $(+1,0)$ or $(+2,0)$ harmonic as the MLH. We investigated whether there is a clean correlation among the BBH system parameters and which harmonic is the MLH, but unfortunately, we have not found any. The bottom panel of Fig.~\ref{fig:dcs} shows the MLH for each system as a function of the lowest PN order SPA correction $\langle \delta \Psi_{3}\rangle$ (which is actually $\mathscr{b}$) and the amplitude weighted phase difference defined as
\begin{equation}
    \label{eq:Ipsi}
    I_{\psi} = \int_{f_{\rm low}}^{f_{\rm high}} df {\cal{A}}_{L}(f) \delta \Psi_{L}(f)\,.
\end{equation}
While a correlation is present, it is not clean, as multiple populations can overlap one another. If there is a correlation between the BBH system parameters and the MLH, it is either higher dimensional (dependent on more than two of the BBH parameters) or depends on a complicated combination of the BBH parameters. Regardless, it is clear from this analysis that the most important harmonics are the $(+1,0)$ and $(+2,0)$ harmonics.

Once we have identified the MLH for each system, we determine the RSH by setting the cutoff to be $90\%$ of the total likelihood. For all of the $10^{4}$ BBH systems, we find that only three harmonics are necessary to meet this requirement. However, there is significant variation in which three harmonics are necessary. Approximately $52\%$ of the BBH systems have the RSH as either $\{(0,0), (+1,0), (+2,0) \}$ or $\{(+1,0), (+1,+2), (+2,0) \}$, with the latter being the most common. The third most common set of RSH is $\{(+1,0), (+1,+2), (+1,-2) \}$. Further investigation reveals that $99.3\%$ of BBH systems require at least one of $\{(+1,0), (+1,+2), (+2,0) \}$ to be part of their respective RSH. In addition, $97.3\%$ of BBH systems have one of these three harmonics as the MLH. We thus conclude that the most important set of harmonics for dCS gravity are the $\{(+1,0), (+1,+2), (+2,0) \}$ harmonics, since these are the most important to the total likelihood computation, with the mappings between ppE and dCS parameters given by Eqs.~\eqref{eq:b-dcs}-\eqref{eq:a-dcs}.

\vspace{0.2cm}
\begin{center}
\noindent{\emph{b. Likelihood analysis of non-axisymmetric \\quadrupoles}}
\end{center}
\vspace{0.2cm}

Let us now repeat the likelihood analysis but for the case of non-axisymmetric quadrupoles. We here only perform the analysis for $10^{3}$ systems, since the trends are much clearer, as we will show. We randomly select parameters $q_{1,2} \in [10^{-3}, 10^{-1}]$, $\mathscr {a}_{1,2} \in [0,2\pi]$, $\beta_{0} \in [0,\pi/2]$, $\omega_{0} \in [0,2\pi]$, and $\chi_{Q} \in [10^{-1}, 10]$ all with a flat prior. Varying the total mass and mass ratio does not impact our results since the dependence of the non-axisymmetric corrections on these has been absorbed into $\chi_{Q}$. 

\begin{figure*}
    \centering
    \includegraphics[width=\textwidth, trim={6cm 0cm 6cm 2cm}, clip]{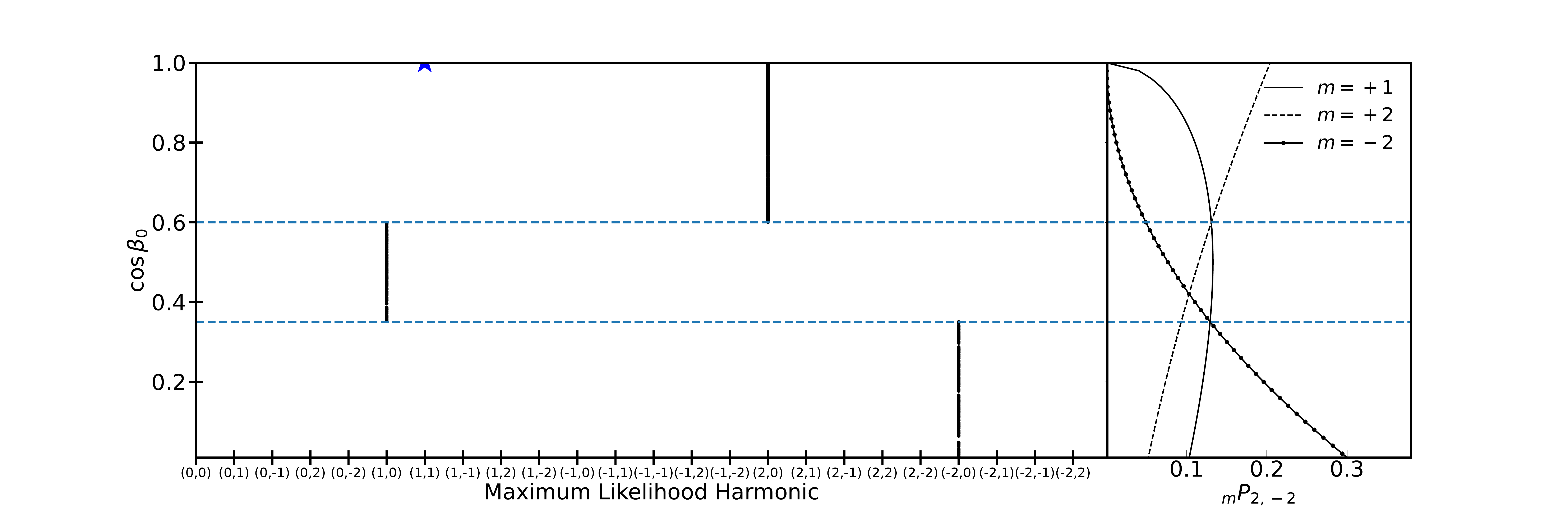}
    \caption{\label{fig:quad} Left: MLH for $10^{3}$ systems with non-axisymmetric quadrupole corrections for different $\cos\beta_{0}$. Observe that there is a clear trend with the initial inclination angle $\beta_{0}$ that determines which harmonic is the MLH. There is one outlier to this trend, indicated by the blue star. Right: SALPs ${_{m}}P_{2,-2}(\beta_{0})$ for different $\cos\beta_0$. Observe that the MLH is determined by which of these has the largest value. The blue dashed line show the separatrix when one of the SALPs becomes larger than the others, indicating a change in the MLH.}
\end{figure*}

Figure~\ref{fig:quad} shows the results for the MLH for each of the $10^{3}$ systems studied. Note that in this scenario, it is possible to have $n\pm1$ harmonics, since the axial modes break the symmetry associated with only have even-$n$ harmonics. A very clear correlation is present between the MLH and the initial inclination angle $\beta_{0}$. This is further elucidated by the right panel of Fig.~\ref{fig:quad}, which displays the behavior of the SALPs ${_{m'}}P_{2,-2}(\beta_{0})$. The MLH is thus determined by which of the SALPs has the largest value. All but one of the systems follow this trend. For the one exception, the $(+1,+1)$ harmonic is the MLH. This is due to the specifics of the initial orientation forcing the $(+2,0)$ harmonic to be suppressed.

\begin{figure*}
    \centering
    \includegraphics[width=\textwidth, trim={5cm 1cm 6cm 2cm}, clip]{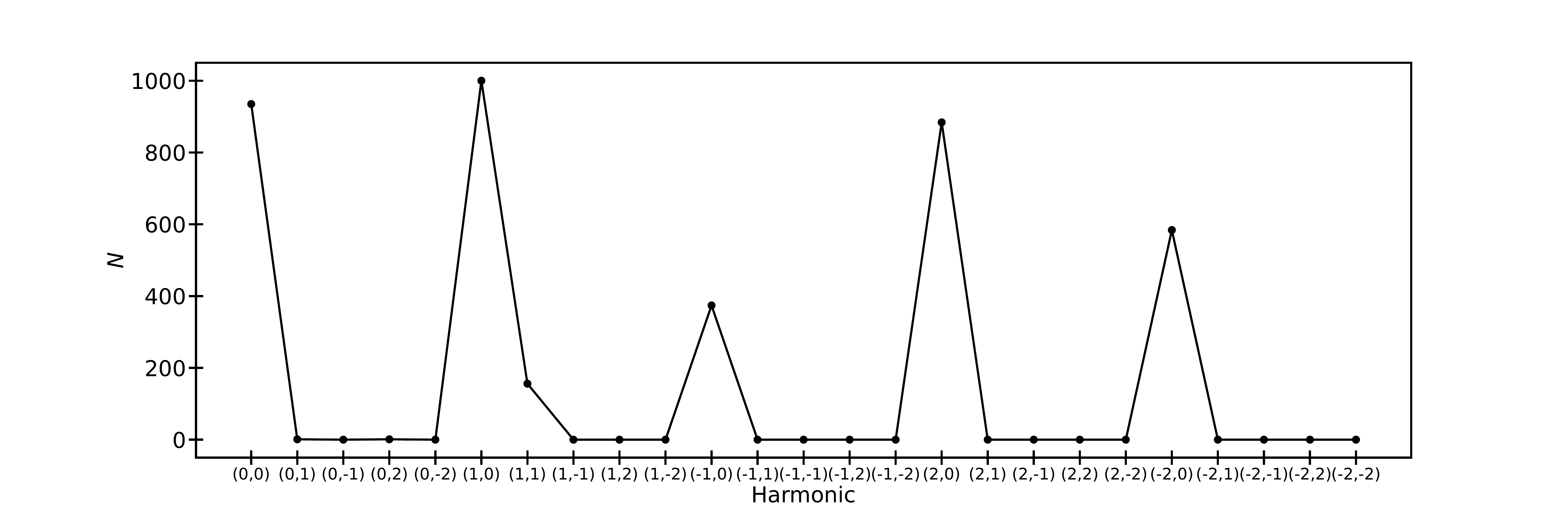}
    \caption{\label{fig:quad-rsh}~Number of systems $N$ with non-axisymmetric quadrupoles whose RSH contains each harmonic. All of the systems have the $(m',n) = (+1,0)$ harmonic as part of their RSH. The next three most common are the $(0,0)$, $(+2,0)$, and $(-2,0)$ harmonics. }
\end{figure*}

With this analysis in hand, let us now discuss the RSH for each system that is necessary to obtain $90\%$ of the total likelihood. For all of the systems, four harmonics are necessary to meet the requirement. Figure~\ref{fig:quad-rsh} displays the number of systems for which each $(m',n)$ harmonic belongs to the RSH. All of the systems' RSH contains the $(+1,0)$ harmonics, while the next three most likely are the $(0,0)$, $(+2,0)$, and $(-2,0)$ harmonics, in descending order respectively. These four harmonics are the subset that we chose to parameterized with ppE deformations, with the mappings given in Eqs.~\eqref{eq:phase-quad-ppE}-\eqref{eq:amp20-quad-ppE}.

\subsubsection{Overlap Considerations}
\label{sec:overlaps}

Once we have determined what the MLH and RSH are for each case, and selected which harmonics to parameterize with ppE deformations, how do we know that we have selected the right set of harmonics? This question can be answered in many ways, depending on what we mean by ``the right set.'' Here, let us answer this question by requiring that the truncated waveform be \textit{faithful} to the full waveform.  Faithfulness then requires that we compute the overlap between a precessing ppE waveform (truncated at a given set of harmonics) and a precessing non-GR waveform that contains deformations in all of the harmonics. 

The overlap is a data analysis statistic used to determine the minimum accuracy for the detection of signals. Mathematically, this quantity is simply given by
\begin{equation}
    {\cal{O}} = \frac{\langle h_{\rm ppE} | h_{\rm non-GR}\rangle}{\sqrt{ \langle h_{\rm ppE} | h_{\rm ppE}\rangle \langle h_{\rm non-GR} | h_{\rm non-GR}\rangle}}\,,
\end{equation}
with its normalization chosen such that ${\cal{O}} \in [0,1]$. 
Commonly, one requires that faithful waveforms be such that ${\cal{O}} > 0.97$, so that no more than 10\% of signals are non-detected due to inaccuracies in the waveform model~\cite{Buonanno:2009zt}. 

\begin{figure}[hbt]
    \centering
    \includegraphics[width=\columnwidth, trim={1cm 0cm 2cm 0cm}, clip]{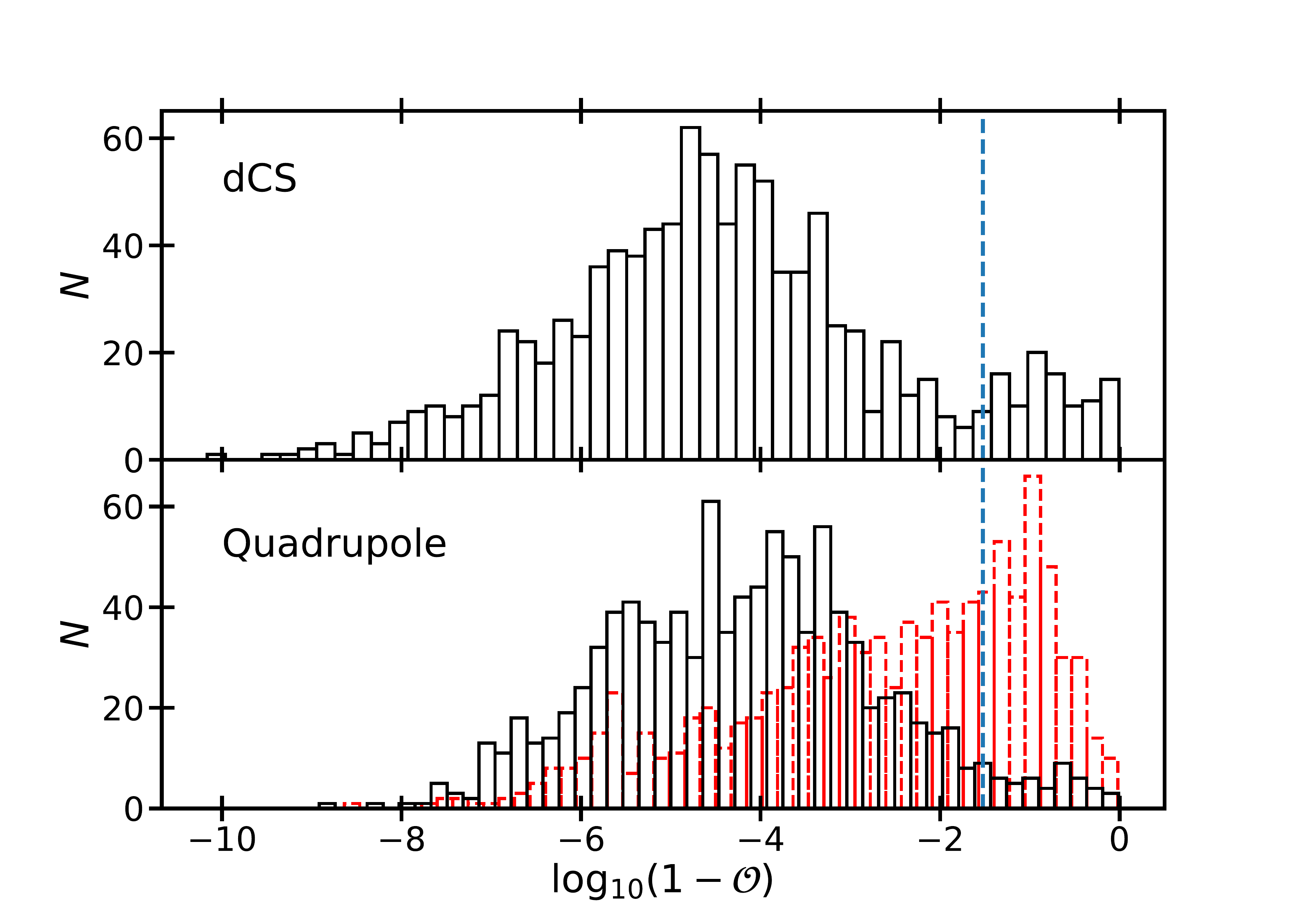}
    \caption{\label{fig:match} Mismatch between the precessing ppE waveforms and non-GR waveforms containing corrections in all harmonics for dCS gravity (top) and non-axisymmetric quadrupoles (bottom), with $N$ the number of systems. The vertical dashed line corresponds to a mismatch of 0.03, or overlap of 0.97, which is a commonly used as a requirement for waveform accuracy. For dCS gravity, 89.6\% of systems have mismatches below this requirement, while for non-axisymmetric quadrupoles 66.7\% or 95.2\% are below this if one include four harmonics (red histogram) or five harmonics (black histogram) in the ppE waveform, respectively.}
\end{figure}

Figure~\ref{fig:match} shows a histogram of the overlap in dCS gravity (top panel) and for non-axisymmetric quadrupoles (bottom panel) computed over $10^3$ systems as a function of the mismatch $1-{\cal{O}}$. For the dCS gravity case, we take $\bar{\zeta}_{2} = 10^{-2}$ for each system to avoid issues related to the equal mass limit (see~\cite{Loutrel:2022tbk}). For the non-axisymemtric quadrupole scenario, we allow $(\cq_{1}, \cq_{2})$ to vary, but still require these to be less than $10^{-1}$. The vertical dashed line provides the detection requirement ${\cal{O}} > 0.97$, and systems to the left of this line meet the requirement. For dCS gravity, $89.6\%$ of systems meet this requirement, while for non-axisymmetric quadrupoles (red dashed histogram), $66.7\%$ of systems meet this requirement. 

For the quadrupole case, the low percentage of systems that meet the ${\cal{O}}>0.97$ requirement is due to the fact that one harmonic is missing from the set of ppE corrections given in Eq.~\eqref{eq:phase-quad-ppE}. While the MLH for the quadrupole case only contains four harmonics for all of the $10^3$ systems studied, Fig.~\ref{fig:quad-rsh} shows that the $(-1,0)$ harmonic is relevant to approximately $40\%$ of the systems. As a result, one might propose adding an additional harmonic for \textit{detection} of such effects (though not necessarily for constraints or tests). The ppE amplitude parameter for this harmonic is
\begin{align}
    \mathscr{a}^{\rm ppE}_{(-1,0)} &= \mathscr{q}_{1} \sin\Delta \left(2 \cot\beta_{0} + \csc\beta_{0}\right) 
    \nn \\
    &- \mathscr{q}_{2} \left(1 + \frac{1}{2} \sec\beta_{0}\right)\,,
    \\
    a_{(-1,0)}^{\rm ppE} &= 0\,.
\end{align}
while the phase correction is still given by Eq.~\eqref{eq:amp20-quad-ppE}. After including this fifth harmonic, the new overlap results are displayed by the black histogram in the bottom panel of Fig.~\ref{fig:match}. The number of systems that now meet the $O>0.97$ requirement is $95.2\%$.

Note that, both in dCS gravity and in the non-axisymmetric quadrupole case, for a significant portion of systems the overlap is much higher, $1-{\cal O}\lesssim 10^{-5}$. As a further useful rule of thumb, two waveforms are considered indistinguishable for parameter estimation purposes if $1-{\cal{O}}\lesssim 1/(2{\rm SNR}^2)$, where SNR is the signal-to-noise ratio~\cite{Lindblom:2008cm}. This implies that the truncated ppE waveforms are indistinguishable from the exact ones as long as ${\rm SNR}\approx45$ for these systems.

While we provide the overlap results here as a means of testing whether our analysis is consistent, it is important to point out that it does not provide any details regarding the intricacies of parameter estimation using the precessing ppE framework. The overlap is a measure of how faithful a model is to an injection, and is related to the number of systems that would be lost in an actual detection search with said model~\cite{Buonanno:2009zt}. In a hypothetical example where GR is not the correct theory of nature (or at least vacuum GR is not the correct model of coalescing BBHs), the overlap informs us that the precessing ppE waveform developed here is suitable for actual searches. 

In a more realistic scenario that studies our ability to carry out tests of GR, one would perform a null test, wherein nature is described by GR and we would use the precessing ppE waveforms developed here to place upper bounds on the non-GR parameters. In such a scenario, the overlap provides no information regarding the results of parameter estimation to obtain these constraints. Indeed, for a test of GR it may well be that much fewer harmonics are needed to obtain a conservative bound (in a similar fashion to needing just one ppE phase deformation at leading PN order to carry out tests without precessing waveforms~\cite{Perkins:2022fhr}). The number of harmonics needed in tests of GR with precessing signals can only be properly studied through Bayesian inference of the posterior probability distribution on the model parameters, which is outside the scope of this paper. 

\section{\label{sec:disc}Discussion}

We have here developed a new ppE framework specifically designed for precessing binaries. The new framework is characterized by $2n$ amplitude parameters $(\mathscr{b}^{\rm ppE}_{\slashed{K}}, \mathscr{a}^{\rm ppE}_{\slashed{K}})$, and $2n$ exponent parameters $(b^{\rm ppE}_{\slashed{K}}, a^{\rm ppE}_{\slashed{K}})$ where $n$ is the number of harmonics used. The need for these harmonics is owed to the richness of precessional dynamics, and is in direct contrast to the simplicity of the original ppE framework. If one employs the likelihood function to determine how many harmonics are needed, we find that $n=3$ for dCS gravity and $n=4$ for non-axisymmetric quadrupoles (and $n=5$ for a more faithful model). 

In spite of these many ppE parameters, however, not all need to be included in the model parameter vector when carrying out tests of GR. This is because half of them are ppE exponent parameters that are fixed in each test. Others are not independent of each other because they will all be proportional to the coupling constant(s) of the modified theory considered. Indeed, in the non-precessing case multiple ppE parameters need to be included to capture high PN order terms, but only one of them (the dominant one) has a physical meaning that maps to the coupling constant of the modified theory, while all others can be marginalized over~\cite{Perkins:2022fhr}. A similar situation may arise in the precessing case considered here, but a deep investigation of these issues is left to future work.  

There are other open questions that the analysis performed here raises. First, are multiple harmonics really needed to place stringent constraints on non-GR coupling parameters with precessing binaries? While the simplified likelihood analysis performed here suggests the answer is yes, this does not take into account covariances, degeneracies, and other intricacies of proper parameter estimation. The overlap analysis informs us that having more harmonics results in a more faithful model, but in the context of non-precessing binaries, more faithful models do not necessarily allow significantly more stringent constraints~\cite{Perkins:2021mhb,Perkins:2022fhr}. Unfortunately, this cannot be answered with the analysis carried out here, and one would have to study the properties of the posterior probability distribution to determine whether this still holds true with precessing binaries.

Second, why are more harmonics necessary in the non-axisymmetric quadrupole case compared to the case of dCS gravity? A likely explanation is that in the former, all of the precessing corrections are degenerate (i.e. they appear at the same PN order), while in the latter, this does not happen. This question also plays into the first question, in that it can also only be answered in a rigorous parameter estimation study using Bayesian inference. Such a study goes outside the scope of this paper, and we leave it for both scenarios (dCS and non-axisymmetric quadrupoles) to future work.

The framework we have developed here only used two example scenarios, dCS gravity and generic multipole moments, as motivation. Work needs to be done to ensure that other beyond-GR scenarios can be adequately mapped to this framework, or to extend the framework to to accommodate more scenarios. In addition, the original ppE framework showed its usefulness long before the first detection~\cite{Cornish:2011ys,Sampson:2013jpa,Sampson:2013lpa,Sampson:2013wia}. Similar studies will need to be done to ensure that the new framework is just as useful for real data analysis scenarios. In this respect, the new ppE parameters for precessing binaries are a blessing and a curse. On the one hand they will make model-agnostic parameter estimation more demanding but, on the other hand, they can greatly help in disentangling degeneracies among different binary's intrinsic parameters and beyond-GR terms.
In particular, it would be instructive to show that the beyond-GR waveforms developed here and in~\cite{Loutrel:2022ant,Loutrel:2022tbk} are capable of breaking parameter degeneracies and placing stringent constraints on the relevant beyond-GR parameters. Regardless, the future of probing fundamental physics with GWs from precessing binaries appears rich.

\acknowledgements
N.L. and P.P. acknowledges financial support provided under the European Union's H2020 ERC, Starting Grant agreement no.~DarkGRA--757480. We also acknowledge support under the MIUR PRIN and FARE programmes (GW-NEXT, CUP: B84I20000100001), and from the Amaldi Research Center funded by the MIUR program ``Dipartimento di Eccellenza'' (CUP: B81I18001170001).
N.Y. acknowledges support from the Simons Foundation through Award number 896696. 
This work is partially supported by the PRIN Grant 2020KR4KN2 ``String Theory as a bridge between Gauge Theories and Quantum Gravity''.

\appendix
\section{Analytic Expressions for Precession Quantities}
In this appendix, we provide some of the expressions for various precession quantities that appear in the discussion in Sec.~\ref{sec:examps}. Many of the analytic expressions are exceedingly lengthy, and we do not provide all of them here. We include those that are immediately relevant to the examples in Sec.~\ref{sec:examps}, and point the reader to~\cite{Loutrel:2022tbk, Loutrel:2022ant} for further reading.
\subsection{Precession Dynamics in dCS Gravity}
\label{app:dcs}

The spin precession equations in dCS gravity take the form
\begin{align}
    \label{eq:prec}
    \dot{\vec{S}}_{1} &= \vec{\Omega} \times \vec{S}_{1}
\end{align}
with
\begin{align}
    \vec{\Omega} &= \vec{\Omega}_{\rm SO} + \vec{\Omega}_{\rm SS} + \vec{\Omega}_{\rm QM}\,,
    \\
    \vec{\Omega}_{\rm SO} &= \frac{\eta}{M} u^{5} \left(2 + \frac{3}{2} q\right) \hat{L}\,,
    \\
    \vec{\Omega}_{\rm SS} &= \frac{1}{2M^{3}} u^{6} \left(1 + \frac{25}{16} \frac{\xi}{m_{1}^{2} m_{2}^{2}}\right) \left[ \vec{S}_{2} - 3 \left(\hat{L} \cdot \vec{S}_{2}\right) \hat{L}\right]\,,
    \\
    \vec{\Omega}_{\rm QM} &= - \frac{3}{2M^{3}} q v^{6} \left(1 - \frac{201}{112} \frac{\xi}{m_{1}^{4}}\right) \left(\hat{L} \cdot \vec{S}_{1}\right) \hat{L}\,.
\end{align}
The precession equation for the second body can be found by taking $1\leftrightarrow2$ in the above expressions, while the equation for the orbital angular momentum $\vec{L}$ can be found from $\dot{\vec{L}} = - \dot{\vec{S}}_{1} - \dot{\vec{S}}_{2}$ due to conservation of the total angular momentum on the precession timescale. In a co-precessing reference frame, the dynamics reduce to solving for the total spin magnitude $S^{2}$, which obeys
\begin{equation}
    \left(\frac{dS^{2}}{dt}\right)^{2} = A^{2} \left(S^{2} - S_{+}^{2}\right) \left(S^{2} - S_{-}^{2}\right) \left(S^{2} - S_{3}^{2}\right)\,,
\end{equation}
where $[S_{\pm}^{2}, S_{3}^{2}]$ are the roots of the polynomial on the right hand side. Each of the roots admits a PN expansion of the form
\begin{align}
S_{\pm}^{2} &= \sum_{n=0} s_{\pm}^{(n)} u^{n} + \bar{\zeta}_{2} \delta s_{\pm}^{(0)}\,,
\\
S_{3}^{2} &= \sum_{n=0} s_{3}^{(n)} u^{n-2} + \bar{\zeta}_{2} \delta s_{3}^{(2)}\,.
\end{align}
The full expressions for the PN coeffiencts $[s_{\pm}^{(n)}, s_{3}^{(n)}]$ and the dCS corrections $[\delta s_{\pm}^{(0)}, \delta s_{3}^{(2)}]$ are exceedingly lenghty, and we simply point the reader to Appendix B of~\cite{Loutrel:2022tbk}. 

The nutation phase $\psi$ obeys the equation
\begin{align}
\frac{d\psi}{dt} &= \frac{A}{2} \sqrt{S_{+}^{2} - S_{-}^{2}}\,,
\end{align}
and evolves under raditation reaction according to Eq.~\eqref{eq:psi-dcs}. Up to 1PN order, the PN coefficients in GR are
\begin{align}
    \label{eq:psi1}
    \psi_{1} &= \frac{3}{4 g_{0} s_{3}^{(0)}} \left[2 g_{1} s_{3}^{(0)} - g_{0} \left(s_{3}^{(1)} + 2 s_{3}^{(0)} \chi_{c}\right)\right]\,,
    \\
    \label{eq:psi2}
    \psi_{2} &= \frac{3}{8g_{0} \left(s_{3}^{(0)}\right)^{2}} \left\{8 g_{2} \left(s_{3}^{(0)}\right)^{2} - 4 g_{1} s_{3}^{(0)} \left(s_{3}^{(1)} + 2 s_{3}^{(0)} \chi_{c}\right) 
    \right.
    \nn \\
    &\left.
    - g_{0} \left[\left(s_{3}^{(1)}\right)^{2} + 4 s_{3}^{(0)} s_{3}^{(2)} - 4 s_{3}^{(0)} s_{+}^{(0)} - 4 s_{3}^{(0)} s_{3}^{(1)} \chi_{c}\right]\right\}\,,
\end{align}
where the $g_{n}$ coefficients are the PN coefficients of $du/dt$ given in Appendix A of~\cite{Chatziioannou:2017tdw}. The dCS correction to $\psi$ is given by
\begin{align}
    \label{eq:dpsi2}
    \delta \psi_{2} &= \frac{\delta y_{0} M^{4} (1-q)^{2} \eta^{2}}{2\sqrt{6 y_{0}} q s_{3}^{(0)}} 
    \nonumber \\
    &+ \frac{3}{s_{3}^{(0)}} \left[\frac{25}{16} c_{1}^{2} q^{3} - \frac{25}{16} \frac{c_{1} M^{2} q^{4} (3+q) \chi_{c}}{(1+q)^{2}} 
    \right.
    \nonumber \\
    &\left.
    - \frac{1}{2} \left(\delta s_{+}^{(0)} + \delta s_{-}^{(0)}\right) - \left(s_{3}^{(0)} - M^{4} \eta\right) \chi_{e,1}\right]\,.
\end{align}
where
\begin{align}
    y_{0} &= \frac{54 q^{2} \Delta_{1} \Delta_{2}}{M^{8}(1-q)^{8}(1+q)^{4} \eta^{4}}\,,
    \\
    \delta y_{0} &= \frac{225c_{1}q^{5}}{8M^{8}(1-q)^{6}(1+q)^{5}\eta^{4}} 
    \nonumber \\
    &\times
    \Big\{(1-q)^{3}(1+q)^{2}\left(S_{2}^{2}-S_{1}^{2}\right)\left[c_{1}(1+q) - q M^{2} \chi_{c}\right] 
    \nonumber \\
    &- \left[2 c_{1}^{2} q (1+q)^{2} - (1-q^{2})^{2} \left(S_{1}^{2} + S_{2}^{2}\right) 
    \right.
    \nonumber \\
    &\left.
    - 2 c_{1} M^{2} q (1+q)^{2} \chi_{c} + 2 M^{4} q^{2} \chi_{c}^{2}\right]
    \nonumber \\
    &\times \left[c_{1}(1+q)^{2} - q (3+q) M^{2} \chi_{c}\right] \Big\}
\end{align}
with
\begin{align}
    \Delta_{1} &= c_{1}^2 (1 + q)^2 - (-1 + q^2)^2 S_{1}^2 
    \nonumber \\
    &- 2 c_{1} M^2 q (1 + q) \chi_{\rm c} +M^4 q^2 \chi_{\rm c}^2\,,
    \\
    \Delta_{2} &= c_{1}^2 q^2 (1 + q)^2 - (-1 + q^2)^2 S_{2}^2 
    \nonumber \\
    &- 2 c_{1} M^2 q^2 (1 + q) \chi_{\rm c} + M^4 q^2 \chi_{\rm c}^2\,.
\end{align}
The dCS modification of the SPA phase of the waveform is given in Eqs.~\eqref{eq:spa-dcs-1}-\eqref{eq:spa-dcs-2}, where
\begin{align}
    \label{eq:db3}
    \langle \delta \beta_{3}^{\rm dCS} \rangle_{\psi} &= \frac{625}{192} \frac{q^{2}}{M^{2}} \frac{(1-q)^{2}}{(1+q)^{2}} \left[c_{1} (1 + q) - q M^{2} \chi_{c} \right]\,,
    \\
    \label{eq:ds4}
    \langle \delta \sigma_{4}^{\rm dCS} \rangle_{\psi} &= -\frac{25}{2304} \frac{q^{2}}{M^{4}(1+q)} \left[c_{1}(1+q) - M^{2} q \chi_{c}\right] 
    \nn \\
    &\times \left[2c_{1}(1+q)^{2} + M^{2} \left(719 - 1442 q + 719 q^{2}\right)\chi_{c} \right]\,,
\end{align}
and
\begin{widetext}
\begin{align}
    \label{eq:dC-avg}
    \langle \delta C\rangle_{\psi} &= \frac{5}{344064} \chi_{1} \frac{(1+q)^{2}}{q^{3}} \left(17815+20311q^{3}\right) + \frac{5}{344064} \chi_{2} \frac{(1+q)^{2}}{q^{3}} \left(20311 + 17815 q^{3}\right) 
    \nonumber \\
    &-\frac{5(1+q)^{6}}{688128 M^{2}(1-q)^{2}q^{4}}\left[18 c_{1}^{2} \left(6537+11410 q^{2} + 6537 q^{4}\right) + 17815 (1-q)^{2} q \left(s_{+}^{(0)} + s_{-}^{(0)}\right)\right]
    \nonumber \\
    &+ \frac{15c_{1}\chi_{c}(1+q)^{6}}{57343M^{2}(1-q)^{2}q^{4}}\left(6537-6537q+12242q^{2}-6537q^{3}+6537q^{4}\right)
    \nonumber \\
    &-\frac{15\chi_{c}^{2}(1+q)^{4}}{114688(1-q)^{2}q^{4}}\left(6537+11410q^{3}+6537q^{6}\right)\,.
\end{align}
\end{widetext}
The secular evolution equations for the precession angle $\alpha$ and Thomas phase $\epsilon$ take the following form in a PN expansion
\begin{align}
    \frac{d\alpha_{-1}}{dt} &= J(u) \left[ \sum_{n=6} \Omega_{z,n}^{(-1)} u^{n} + \bar{\zeta}_{2} \delta \Omega_{z,6}^{(-1)} u^{6}\right]\,,
    \\
    \frac{d\epsilon_{-1}}{dt} &= \sum_{n=5} \Omega_{T,n}^{(-1)} u^{n} + \bar{\zeta}_{2} \delta \Omega_{T,6}^{(-1)} u^{5}\,,
\end{align}
where the full expressions for $[\Omega_{z,n}^{(-1)},\Omega_{T,n}^{(-1)}]$ are given in Appendix D of~\cite{Loutrel:2022tbk}. The solutions to these equations take the forms of Eq.~\eqref{eq:alpha-dcs} for $\alpha_{-1}(u)$ and Eq.~\eqref{eq:eps-dcs} for $\epsilon_{-1}(u)$, where the relevant coefficients are
\begin{align}
    \label{eq:Phiz-3}
    \alpha_{-3}^{(-1)} &= \frac{3M \Omega_{z,6}^{(-1)}}{a_{0}}\,,
    \\
    \label{eq:Phiz-2}
    \alpha_{-2}^{(-1)} &= \frac{3M \Omega_{z,7}^{(-1)}}{a_{0}}\,,
    \\
    \label{eq:dPhiz-3}
    \delta \alpha_{-3}^{(-1)} &= \frac{3M \delta \Omega_{z,6}^{(-1)}}{a_{0}}\,.
\end{align}
and
\begin{align}
    \label{eq:PhiT-0}
    \epsilon_{-3}^{(-1)} &= -\frac{M}{a_{0}} \Omega_{T,5}^{(-1)}\,,
    \\
    \label{eq:PhiT-1}
    \epsilon_{-2}^{(-1)} &= - \frac{3M}{2a_{0}} \Omega_{T,6}^{(-1)}\,,
    \\
    \label{eq:dPhiT-0}
    \delta \epsilon_{-3}^{(-1)} &= - \frac{M}{a_{0}} \delta \Omega_{T,5}^{(-1)}\,,
\end{align}
where $a_{0} = 96\eta/5$.

\subsection{Precession Dynamics for Non-Axisymmetric Compact Objects}
\label{app:quad}

To leading PN order, non-axisymmetry is encoded in the mass quadrupole moment of a compact object, specifically in the $m = \pm 1, \pm 2$ modes. Thus, for binary systems, the leading order deviations will occur in the quadrupole-monopole interaction. Using the method of osculating orbits, the secular evolution equations for the orientation angles $[\beta, \alpha, \omega]$ become~\cite{Loutrel:2022ant}
\begin{align}
\frac{d\beta}{dt} &= \sqrt{6} \Omega_{Q} \Big\{ \cos\beta \left(\mathscr{r}_{1} \sin\alpha + \mathscr{i}_{1} \cos\alpha \right) 
\nn \\
&- \sin\beta \left[\mathscr{r}_{2} \sin(2\alpha) + \mathscr{i}_{2} \cos(2\alpha)\right] \Big\}\,,
\\
\label{eq:alpha-dot}
\frac{d\alpha}{dt} &= \Omega_{Q} \Big\{\cos\beta \left[3 - \sqrt{6} \mathscr{r}_{2} \cos(2\alpha) + \sqrt{6} \mathscr{i}_{2} \sin(2\alpha) \right] 
\nn \\
&+ \sqrt{6} \cos(2\beta) \left(\mathscr{r}_{1} \cos\alpha - \mathscr{i}_{1} \sin\alpha \right) \Big\}
\\
\frac{d\omega}{dt} &= \frac{\Omega_{Q}}{4} \Big\{-9 - 15 \cos(2\beta) 
\nn \\
&+ \sqrt{6} \left[6 - 10 \cos(2\beta)\right] \cot\beta \left(\mathscr{r}_{1} \cos\alpha - \mathscr{i}_{1} \sin\alpha \right) 
\nn \\
&+ \sqrt{6} \left[1 - 5 \cos(2\beta)\right] \left[\mathscr{i}_{2} \sin(2\alpha) - \mathscr{r}_{2} \cos(2\alpha) \right] \Big\}
\end{align}
where
\begin{align}
    \Omega_{Q} &= \left(\frac{\pi}{5}\right)^{1/2} \frac{Q_{0} (1-e^{2})^{3/2}}{\nu M^{1/2} p^{7/2}}
    \\
    Q_{m}/Q_{0} &= \mathscr{r}_{m} + i \mathscr{i}_{m}
\end{align}
with $[e,p]$ the osculating eccentricity and semi-latus rectum, $Q_{m}$ the mass quadrupole coefficients in a spherical harmonic basis, and $[\nu, M]$ the symmetric mass ratio and total mass of the binary. Defining
\begin{equation}
    \cq_{m} = \sqrt{\frac{2}{3} \left(\mathscr{r}_{m}^{2} + \mathscr{i}_{m}^{2}\right)}\,, \qquad \ca_{m} = \frac{1}{m} \tan^{-1} \left(\mathscr{i}_{m}/\mathscr{r}_{m}\right)\,,
\end{equation}
reference~\cite{Loutrel:2022ant} showed that exact solution can be found when $\cq_{1} = 0$ and $\cq_{2} \in [0,1)$. The exact solutions are
\begin{align}
    \label{eq:alpha-pol}
    \alpha_{2} &= -\frac{\pi}{2} - \ca_{2} + \tan^{-1}\left[\sqrt{\frac{1 + \cq_{2}}{1 - \cq_{2}}} \tan \psi_{2} \right]\,,
    \\
    \beta_{2} &= \sin^{-1}\left[ \sin\beta_{0} \sqrt{\frac{1 + \cq_{2} \cos(2\psi_{2})}{1 + \cq_{2}}}\right]
    \\
    \omega_{2} &= \omega_{0} + \frac{\sec\beta_{0}}{4 \sqrt{1 - \cq_{2}^{2}}} \left[\frac{c_{2}}{\cq_{2}} {\rm F} \left(\psi_{2} \bigg| \frac{2b}{b - a_{-}}\right) 
    \right.
    \nn \\
    &\left.
    - 4 (1 + \cq_{2}) \Pi\left(\frac{2\cq_{2}}{1-\cq_{2}} ; \psi_{2} \bigg| \frac{2b}{b-a_{-}}\right) \right]
\end{align}
where $[{\rm F}, \Pi]$ are the elliptic integrals of the first and third kind respectively, and
\begin{align}
    \label{eq:a-minus}
    a_{-} &= 1 - \cq_{2} - \sin^{2}\beta_{0}\,,
    \\
    b &= -\cq_{2} \sin^{2}\beta_{0}\,,
    \\
    c_{2} &= \cq_{2} \left(1 + 5 \cq_{2}\right) - 5 \cq_{2} \left(1 + \cq_{2}\right) \cos(2\beta_{0})\,.
\end{align}
The phase variable $\psi_{2}$ evolves on the radiation reaction timescale, with the analytic solution
\begin{equation}
    \label{eq:psi2-pol}
    \psi_{2} = {\rm am} \left[{\rm F} \left(\psi_{c} \bigg| \frac{2b}{b+a_{+}}\right) - \psi(u) \bigg| \frac{2b}{b+a_{+}}\right]\,,
\end{equation}
where ${\rm am}(x|n)$ is the Jacobi amplitude function and $a_{+}$ is given by making the replacement $\cq_{2} \rightarrow - \cq_{2}$ in Eq.~\eqref{eq:a-minus}. More general solutions that allow $\cq_{1} > 0$ were also found by perturbing around these, specifically
\begin{align}
    \label{eq:alpha-full}
    \alpha &= \alpha_{2}(\psi_{2}) + \cq_{1} \delta \alpha_{(1)}(\psi_{2}) + {\cal{O}}(\cq_{1}^{2})\,,
    \\
    \label{eq:beta-full}
    \beta &= \beta_{2}(\psi_{2}) + \cq_{1} \delta \beta_{(1)}(\psi_{2}) + {\cal{O}}(\cq_{1}^{2})\,,
    \\
    \label{eq:omega-full}
    \omega &= \omega_{2}(\psi_{2}) + \cq_{1} \delta \omega_{(1)}(\psi_{2}) + {\cal{O}}(\cq_{1}^{2})\,.
\end{align}
We do not provide the correction functions $[\delta \alpha_{(1)}, \delta\beta_{(1)}, \delta \omega_{(1)}]$ here since they are lengthy, and can be read off of Eqs.~(94)-(99) in~\cite{Loutrel:2022ant}. Likewise, $\psi_{2}$ is also corrected at linear order in $\cq_{1}$, with the relevant correction function given in Eq.~(130) of~\cite{Loutrel:2022ant}. To obtain the relevant expression in Sec.~\ref{sec:quad}, one simply has to take the limit $\cq_{1} \ll 1 \gg \cq_{2}$ of the expressions in Eqs.~\eqref{eq:alpha-pol}-\eqref{eq:omega-full}.
\bibliographystyle{apsrev4-1}
\bibliography{refs}
\end{document}